\documentclass[usenatbib]{mn2e}

\usepackage{psfig,graphicx}

\def\rpd{\hbox{rad\,d$^{-1}$}}
\def\mrpd{\hbox{mrad\,d$^{-1}$}}
\def\omeq{\hbox{$\Omega_{\rm eq}$}}
\def\dom{\hbox{$d\Omega$}}
\def\Bl{\hbox{$B_{\rm \ell}$}}
\def\vr{\hbox{$v_{\rm r}$}}
\def\chisq{\hbox{$\chi^2$}}
\def\chinu{\hbox{$\chi^2_\nu$}}
\def\msun{\hbox{${\rm M}_{\odot}$}}

\def\rsun{\hbox{${\rm R}_{\odot}$}}
\def\mstar{\hbox{$M_{\star}$}}
\def\rstar{\hbox{$R_{\star}$}}

\def\sn{\hbox{S/N}}

\def\kms{\hbox{km\,s$^{-1}$}}
\def\eps{\hbox{erg\,s$^{-1}$}}
\def\vsini{\hbox{$v\sin i$}}

\def\ptt{\hbox{$10^{-4} I_{\rm c}$}}

\def\arcsec{\hbox{$^{\prime\prime}$}}
\def\degr{\hbox{$^\circ$}}

\def\prot{\hbox{$P_{\rm rot}$}}

\def\logLX{\hbox{$\log L_{\rm X}/L_{\rm bol}$}}
\def\logLb{\hbox{$\log L_{\rm bol}$}}
\def\logRx{\hbox{$\log R_{\rm X}$}}

\begin{document}

\title[Magnetic topologies of early M dwarfs] {Large-scale magnetic topologies of early M dwarfs\thanks{Based 
on observations obtained at the T\'elescope Bernard Lyot (TBL), operated by the Institut National 
des Science de l'Univers of the Centre National de la Recherche Scientifique of France.} }

\makeatletter

\def\newauthor{%
  \end{author@tabular}\par
  \begin{author@tabular}[t]{@{}l@{}}}
\makeatother
 
\author[J.-F.~Donati et al]
{\vspace{1.5mm}
J.-F.~Donati$^1$\thanks{E-mail:
donati@ast.obs-mip.fr (J-FD);
jmorin@ast.obs-mip.fr (JM);
petit@ast.obs-mip.fr (PP);
xavier.delfosse@obs.ujf-grenoble.fr (XD);
thierry.forveille@obs.ujf-grenoble.fr (TF);
auriere@ast.obs-mip.fr (MA);
remi.cabanac@ast.obs-mip.fr (RC);
dintrans@ast.obs-mip.fr (BD);
rfares@ast.obs-mip.fr (RF);
tgastine@ast.obs-mip.fr (TG);
mmj@st-and.ac.uk (MMJ);
lignieres@ast.obs-mip.fr (FL);
fpaletou@ast.obs-mip.fr (FP);
julio.ramirez@obspm.fr (JCRV);
sylvie.theado@ast.obs-mip.fr (ST)}, 
J.~Morin$^1$, P.~Petit$^1$, X.~Delfosse$^2$, T.~Forveille$^2$, M.~Auri\`ere$^1$, \\ 
\vspace{1.7mm}
{\hspace{-1.5mm}\LARGE\rm
R.~Cabanac$^1$, B.~Dintrans$^1$, R.~Fares$^1$, T.~Gastine$^1$, M.M.~Jardine$^3$, F.~Ligni\`eres$^1$, } \\
{\hspace{-1.5mm}\LARGE\rm 
F.~Paletou$^1$, J.C.~Ramirez Velez$^4$, S.~Th\'eado$^1$} \\
$^1$ LATT--UMR 5572, CNRS \& Univ.\ de Toulouse, 14 Av.\ E.~Belin, F--31400 Toulouse, France\\
$^2$ LAOG--UMR~5571, CNRS \& Univ.\ J.~Fourier, 31 rue de la Piscine, F--38041 Grenoble, France\\ 
$^3$ School of Physics and Astronomy, Univ.\ of St~Andrews, St~Andrews, Scotland KY16 9SS, UK \\ 
$^4$ LESIA, Observatoire de Paris-Meudon, F--92195 Meudon, France 
}
\date{\today}
\maketitle
 
\begin{abstract}  
We present here additional results of a spectropolarimetric
survey of a small sample of stars ranging from spectral type M0 to M8 
aimed at investigating observationally how dynamo processes operate 
in stars on both sides of the full convection threshold (spectral type M4).  

The present paper focuses on early M stars (M0--M3), 
i.e.\ above the full convection threshold.
Applying tomographic imaging techniques to time series of rotationally 
modulated circularly polarised profiles collected with the NARVAL 
spectropolarimeter, we determine the rotation period and reconstruct 
the large-scale magnetic topologies of 6 early M dwarfs.  
We find that early-M stars preferentially host large-scale fields with 
dominantly toroidal and non-axisymmetric poloidal configurations, along 
with significant differential rotation (and long-term variability);  
only the lowest-mass star of our subsample is found to host an almost fully 
poloidal, mainly axisymmetric large-scale field ressembling those found in 
mid-M dwarfs.  

This abrupt change in the large-scale magnetic topologies of M dwarfs (occuring 
at spectral type M3) has no related signature on X-ray luminosities (measuring 
the total amount of magnetic flux);  it thus suggests that underlying dynamo 
processes become more efficient at producing large-scale fields (despite 
producing the same flux) at spectral types later than M3.  We suspect that 
this change relates to the rapid decrease in the radiative cores of low-mass 
stars and to the simultaneous sharp increase of the convective turnover times 
(with decreasing stellar mass) that models predict to occur at M3;  it may 
also be (at least partly) responsible for the reduced magnetic braking reported 
for fully-convective stars.  
\end{abstract}


\section{Introduction}

Activity and magnetic fields are ubiquitous to cool stars of all spectral 
types \citep[e.g.,][]{Saar85, Donati97}.  The analogy with the Sun suggests 
that these fields are produced through dynamo mechanisms operating in a thin 
interface layer at the base of the convective envelope, where angular rotation 
gradients are strongest.  Starting from a weak poloidal configuration, 
differential rotation progressively winds the field around the star, building 
a strong toroidal belt at the base of the convective zone;  cyclonic turbulence 
then restores a weak poloidal field (with a polarity opposite to that of the 
initial one) once the toroidal field has grown unstable \citep{Parker55}.  
However, there is still considerable controversy (even for the Sun itself) on 
how and where exactly the field is amplified and on which physical processes 
(differential rotation, meridional circulation) are mainly controlling the 
magnetic cycle \citep[e.g.,][]{Charbonneau05}.  

Observing stars other than the Sun is of obvious interest for this question 
as they provide a direct way of studying how dynamo processes depend on 
fundamental stellar parameters such as mass (and thus convective depth) and 
rotation rate.  In this respect, low-mass fully-convective stars are 
particularly interesting as they host no interface layer (where 
dynamo processes presumably concentrate in the Sun), and nevertheless show 
both strong magnetic fields \citep[e.g.,][]{Johns96} and intense activity 
\citep[e.g.,][]{Delfosse98}.  Many theoretical models were proposed to 
attempt resolving this issue \citep[e.g.,][]{Durney93, Dobler06, Browning08} 
but observations of the large-scale magnetic fields at the surfaces of 
fully-convective stars, and in particular of their poloidal and toroidal 
components, are still rare.  

Thanks to time-resolved spectropolarimetric observations of cool stars, 
we are now able to recover information on how magnetic fields distribute 
at (and emerge from) the surfaces of cool active stars 
\citep[e.g.,][]{Donati03a}.  With the advent of new generation instruments 
optimised in this very purpose \citep[ESPaDOnS at the 3.6m Canada-France-Hawaii 
Telescope and NARVAL at the 2m T\'elescope Bernard Lyot in France, ][]{Donati03c}, 
this technique can now be also applied to faint M dwarfs.  The first attempt 
focussed on the fully-convective rapidly-rotating M4 star V374~Peg and revealed 
that, against all theoretical expectations, fully-convective M dwarfs are 
apparently very efficient at producing strong large-scale mainly-axisymmetric 
poloidal fields \citep{Donati06} despite very small levels of differential 
rotation.  A follow-up study confirmed this point and further demonstrated 
that the magnetic configuration of V374~Peg is apparently stable on timescales 
of $\simeq1$~yr \citep{Morin08}.  

To investigate this issue in more details, we embarked in a spectropolarimetric 
survey of a small sample of M dwarfs, located both above and below the 
full-convection threshold (corresponding to a mass of 0.35~\msun\ and to a 
spectral type of M4).  The first results of this survey, focussing mainly on 
active M4 dwarfs, confirms the results obtained on V374~Peg and demonstrate 
that strong large-scale mainly-axisymmetric poloidal fields are indeed fairly  
common in fully-convective mid-M active dwarfs \citep[][hereafter M08]{Morin08b}.  
In this new paper, we present the results for the 6 early M dwarfs that we observed 
in this survey, namely DT~Vir, DS~Leo, CE~Boo, OT~Ser, GJ~182 and GJ~49, 
with spectral types ranging from M0 to M3 (see Table~\ref{tab:sample}).  
After briefly describing the observations and the magnetic modelling method 
that we use, we detail the results obtained for each star and discuss their 
implication for our understanding of dynamo processes in cool active stars.  

\begin{table*}
\caption{Fundamental parameters of our sample early-M dwarf stars.  
Columns 1 to 6 respectively list the name, the spectral type, the 
mass \mstar\ (derived from the Hipparcos distance and the J, H and K magnitudes 
using the mass-luminosity relations of \citealt{Delfosse00} except for GJ~182; 
for which we used the evolutionary models of \citealt{Baraffe98}, see text), the 
logarithmic bolometric luminosity \logLb\ (derived from the mass and the models of 
\citealt{Baraffe98}), the logarithmic relative X-ray luminosity \logRx\ (i.e., 
\logLX, from \citealt{Kiraga07} or from the NEXXUS data base,  
\citealt{Schmitt04}, or from \citealt{Wood94} for GJ~49), the 
projected rotation velocity (this paper, accuracy $\simeq1$~\kms), the rotation 
period (this paper),  the convective turnover time $\tau_{\rm c}$ (from 
\citealt{Kiraga07}), the effective Rossby number $Ro=\prot/\tau_{\rm c}$, 
the radius \rstar\ (predicted by the theoretical models of \citealt{Baraffe98}) 
and the assumed inclination angle of the rotation axis to the line-of-sight (this 
paper). }
 \begin{tabular}{lccccccccccc}
 \hline
Star & ST & \mstar & \logLb & \logRx & \vsini & \prot & $\tau_{\rm c}$ &  $Ro$ & \rstar & $i$ \\
     &    & (\msun) & (\eps)&        & (\kms) & (d) & (d) & & (\rsun) & (\degr) \\
 \hline
GJ 182           & M0.5 & 0.75 & 32.7 & --3.1 & 10 & 4.35 & 25 & 0.174 & 0.82 & 60 \\ 
DT Vir / GJ 494A & M0.5 & 0.59 & 32.3 & --3.4 & 11 & 2.85 & 31 & 0.092 & 0.53 & 60 \\ 
DS Leo / GJ 410  & M0   & 0.58 & 32.3 & --4.0 &  2 & 14.0 & 32 & 0.438 & 0.52 & 60 \\ 
GJ 49            & M1.5 & 0.57 & 32.3 & $<-4.3$ &  1 & 18.6 & 33 & 0.564 & 0.51 & 45 \\ 
OT Ser / GJ 9520 & M1.5 & 0.55 & 32.2 & --3.4 &  6 & 3.40 & 35 & 0.097 & 0.49 & 45 \\ 
CE Boo / GJ 569A & M2.5 & 0.48 & 32.1 & --3.7 &  1 & 14.7 & 42 & 0.350 & 0.43 & 45 \\ 
 \hline
 \end{tabular}
\label{tab:sample}
\end{table*}

\section{Observations}
\label{sec:obs}

Spectropolarimetric observations of the selected M dwarfs were collected with 
NARVAL and the 2m T\'elescope Bernard Lyot (TBL), between 2007 Jan and 2008 Feb 
(in 3 different runs).  NARVAL is a twin spectropolarimeter copied from ESPaDOnS 
\citep{Donati03c}, yielding full coverage of the optical domain (370 to 1000~nm) 
at a resolving power of about 65~000 in a single exposure.  Each polarisation 
exposure consists of 4 individual subexposures taken in different polarimeter 
configurations and combined together to filter out all spurious polarisation 
signatures at first order \citep[e.g.,][]{Donati97}.

Data reduction was carried out using \textsc{Libre ESpRIT}, a fully automated 
package/pipeline installed at TBL and performing optimal extraction of 
unpolarised (Stokes $I$) and circularly polarised (Stokes $V$) spectra 
as described in \citet{Donati97}.  The peak signal-to-noise ratios (\sn) per
2.6~\kms\ velocity bin that we obtained in the collected spectra range from 100 
to 400 depending on the magnitude and weather conditions. The journal of 
observations for all stars is presented in Tables~\ref{tab:log494} to \ref{tab:log49}.

All spectra are automatically corrected from spectral shifts resulting from
instrumental effects (eg mechanical flexures, temperature or pressure variations) using
telluric lines as a reference.  Though not perfect, this procedure allows spectra to be 
secured with a radial velocity (RV) precision of better than 0.030~\kms\ 
\citep[e.g.,][]{Moutou07}.

\begin{table*}
\caption[]{Journal of observations for DT~Vir=GJ~494A. Columns 1--7 list the UT
date, the heliocentric Julian date, the UT time, the total exposure time, 
the peak signal to noise ratio (per 2.6~\kms\ velocity bin) and the rms noise 
level in the LSD Stokes $V$ profile (relative to the unpolarised continuum level 
and per 1.8~\kms\ velocity bin).  
In col.~8 and 9, we list the rotational cycles (using ephemeris 
HJD$=2,454,100.0+2.85 E$) as well as the longitudinal fields \Bl\ (with error 
bars) and the radial velocities \vr\ (absolute accuracy 0.10~\kms, internal accuracy 
0.03~\kms)  associated to each exposure. } 
\begin{tabular}{rcccccrrc}
\hline
Date & HJD          & UT      & $t_{\rm exp}$ & \sn\ & $\sigma_{\rm LSD}$ & Cycle & \Bl  & \vr \\ 
     & (2,454,000+) & (h:m:s) &   (s)         &      &   (\ptt)           &       &  (G) & (\kms) \\
\hline
2007 Jan 26 & 126.70173 & 04:47:56 & 4$\times800$ & 240 & 3.3 &  9.369 & $-17.6\pm9.3$ & $-13.32$ \\ 
     Jan 27 & 127.67710 & 04:12:22 & 4$\times800$ & 310 & 2.3 &  9.711 &  $-3.8\pm6.7$ & $-13.44$ \\ 
     Jan 28 & 128.69001 & 04:30:50 & 4$\times800$ & 350 & 2.0 & 10.067 &  $66.0\pm5.8$ & $-13.04$ \\ 
     Jan 29 & 129.65374 & 03:38:29 & 4$\times800$ & 340 & 2.1 & 10.405 & $-27.6\pm6.0$ & $-13.29$ \\ 
     Feb 02 & 133.71569 & 05:07:13 & 4$\times800$ & 350 & 2.0 & 11.830 &  $23.3\pm5.7$ & $-13.29$ \\ 
     Feb 03 & 134.69961 & 04:43:57 & 4$\times800$ & 340 & 2.1 & 12.175 &  $49.7\pm6.0$ & $-13.00$ \\ 
     Feb 04 & 135.71225 & 05:02:02 & 4$\times800$ & 360 & 1.9 & 12.531 & $-31.1\pm5.6$ & $-13.32$ \\ 
     Feb 05 & 136.66208 & 03:49:41 & 2$\times800$ & 140 & 5.2 & 12.864 &  $28.0\pm15.5$ & $-13.25$ \\ 
\hline
2007 Dec 28 & 462.71536 & 05:11:18 & 4$\times750$ & 330 & 2.2 & 127.269 & $30.5\pm6.5$ & $-13.64$ \\
     Dec 29 & 463.74605 & 05:55:22 & 4$\times750$ & 360 & 2.0 & 127.630 & $15.5\pm5.7$ & $-13.67$ \\
     Dec 31 & 465.74401 & 05:52:10 & 4$\times700$ & 290 & 2.5 & 128.331 & $29.0\pm7.1$ & $-13.66$ \\
2008 Jan 01 & 466.74583 & 05:54:39 & 4$\times700$ & 360 & 2.0 & 128.683 & $24.5\pm5.7$ & $-13.69$ \\

     Jan 19 & 484.63507 & 03:12:52 & 4$\times600$ & 170 & 4.5 & 134.960 & $26.3\pm12.8$ & $-13.57$ \\
     Jan 20 & 485.61280 & 02:40:40 & 4$\times600$ & 200 & 4.0 & 135.303 & $29.4\pm11.3$ & $-13.77$ \\
     Jan 23 & 488.62610 & 02:59:27 & 4$\times600$ & 250 & 3.1 & 136.360 & $23.6\pm9.0$ & $-13.74$ \\
     Jan 24 & 489.62041 & 02:51:08 & 4$\times600$ & 240 & 2.9 & 136.709 & $ 1.5\pm8.6$ & $-13.65$ \\
     Jan 26 & 491.60391 & 02:27:08 & 4$\times600$ & 230 & 3.3 & 137.405 & $39.8\pm9.5$ & $-13.69$ \\
     Jan 27 & 492.62606 & 02:58:54 & 4$\times600$ & 320 & 2.2 & 137.763 & $-4.5\pm6.2$ & $-13.67$ \\
     Jan 28 & 493.63758 & 03:15:22 & 4$\times600$ & 300 & 2.4 & 138.118 & $ 4.5\pm6.7$ & $-13.56$ \\

     Feb 03 & 499.63751 & 03:14:34 & 4$\times600$ & 280 & 2.7 & 140.224 & $ 8.9\pm7.7$ & $-13.80$ \\
     Feb 05 & 501.64773 & 03:29:03 & 4$\times600$ & 290 & 2.4 & 140.929 & $53.9\pm7.0$ & $-13.59$ \\
     Feb 06 & 502.63636 & 03:12:34 & 4$\times600$ & 320 & 2.2 & 141.276 & $-7.1\pm6.4$ & $-13.75$ \\
     Feb 07 & 503.64038 & 03:18:15 & 4$\times600$ & 290 & 2.4 & 141.628 & $23.2\pm7.1$ & $-13.69$ \\
     Feb 10 & 506.64784 & 03:28:40 & 4$\times600$ & 310 & 2.3 & 142.684 & $ 7.0\pm6.6$ & $-13.65$ \\
     Feb 12 & 508.63939 & 03:16:18 & 4$\times600$ & 300 & 2.4 & 143.382 & $-4.9\pm6.9$ & $-13.82$ \\
     Feb 14 & 510.63942 & 03:16:08 & 4$\times600$ & 230 & 3.2 & 144.084 & $ 7.4\pm9.2$ & $-13.61$ \\
     Feb 15 & 511.65542 & 03:39:04 & 4$\times600$ & 280 & 2.5 & 144.440 & $ 6.7\pm7.3$ & $-13.77$ \\
     Feb 16 & 512.64053 & 03:17:32 & 4$\times600$ & 300 & 2.4 & 144.786 & $-2.0\pm6.7$ & $-13.68$ \\
     Feb 17 & 513.64033 & 03:17:09 & 4$\times600$ & 310 & 2.3 & 145.137 & $-0.1\pm6.6$ & $-13.65$ \\
\hline
\end{tabular}
\label{tab:log494}
\end{table*}

\begin{table*}
\caption[]{Same as Table~\ref{tab:log494} for DS~Leo=GJ~410. 
Rotational cycles are now computed using ephemeris HJD$=2,454,100.0+14.0 E$. } 
\begin{tabular}{rcccccrrc}
\hline
Date & HJD          & UT      & $t_{\rm exp}$ & \sn\ & $\sigma_{\rm LSD}$ & Cycle & \Bl & \vr \\ 
     & (2,454,000+) & (h:m:s) &   (s)         &      &   (\ptt)           &       &  (G) & (\kms) \\
\hline
2007 Jan 26 & 126.65015 & 03:31:14 & 4$\times800$ & 250 & 2.7 & 1.904 & $-33.9\pm4.1$  & $-14.08$ \\ 
     Jan 27 & 127.63491 & 03:09:13 & 4$\times800$ & 270 & 2.3 & 1.974 & $-15.9\pm3.4$  & $-14.05$ \\ 
     Jan 28 & 128.64692 & 03:26:27 & 4$\times800$ & 320 & 2.0 & 2.046 & $-1.3\pm3.0$   & $-14.08$ \\ 
     Jan 29 & 129.61070 & 02:34:13 & 4$\times800$ & 330 & 1.9 & 2.115 & $13.1\pm2.9$   & $-14.06$ \\ 
     Jan 30 & 130.65362 & 03:35:57 & 4$\times800$ & 340 & 1.7 & 2.189 & $29.7\pm2.6$   & $-14.08$ \\ 
     Feb 02 & 133.67041 & 03:59:57 & 4$\times800$ & 340 & 1.8 & 2.405 & $6.3\pm2.8$    & $-14.03$ \\ 
     Feb 03 & 134.65677 & 03:40:15 & 4$\times800$ & 340 & 1.8 & 2.475 & $7.0\pm2.8$    & $-14.01$ \\ 
     Feb 04 & 135.66947 & 03:58:28 & 4$\times800$ & 350 & 1.8 & 2.548 & $3.7\pm2.7$    & $-14.02$ \\ 
     Feb 05 & 136.63031 & 03:02:02 & 4$\times800$ & 200 & 3.2 & 2.617 & $2.4\pm5.1$    & $-14.04$ \\ 
\hline
2007 Dec 28 & 462.67598 & 04:11:04 & 4$\times750$ & 320 & 2.0 & 25.905 & $9.9\pm2.9$   & $-14.09$ \\
     Dec 29 & 463.70487 & 04:52:33 & 4$\times750$ & 340 & 1.8 & 25.979 & $1.5\pm2.7$   & $-14.09$ \\
     Dec 31 & 465.70955 & 04:59:06 & 4$\times600$ & 260 & 2.5 & 26.122 & $8.6\pm3.7$   & $-14.08$ \\
2008 Jan 01 & 466.70912 & 04:58:22 & 4$\times700$ & 350 & 1.8 & 26.194 & $24.5\pm2.7$  & $-14.09$ \\

     Jan 02 & 467.68306 & 04:20:44 & 2$\times600$ & 130 & 5.1 & 26.263 & $13.9\pm7.7$  & $-14.07$ \\
     Jan 03 & 468.70748 & 04:55:48 & 4$\times700$ & 180 & 3.8 & 26.336 & $26.9\pm5.6$  & $-14.05$ \\
     Jan 07 & 472.62578 & 02:57:46 & 4$\times700$ & 160 & 4.3 & 26.616 & $-22.1\pm7.0$ & $-14.06$ \\
     Jan 08 & 473.62319 & 02:53:56 & 4$\times700$ & 260 & 2.4 & 26.687 & $-4.0\pm3.9$  & $-14.11$ \\
     Jan 19 & 484.58050 & 01:51:30 & 4$\times600$ & 180 & 4.0 & 27.470 & $-17.2\pm5.9$ & $-14.04$ \\
     Jan 20 & 485.55637 & 01:16:40 & 4$\times600$ & 190 & 3.7 & 27.540 & $-26.8\pm5.5$ & $-14.05$ \\
     Jan 22 & 487.58395 & 01:56:14 & 4$\times600$ & 230 & 2.8 & 27.685 & $-9.6\pm4.4$  & $-14.04$ \\
     Jan 24 & 489.53279 & 00:42:24 & 4$\times600$ & 250 & 2.6 & 27.824 & $6.4\pm3.8$   & $-14.05$ \\
     Jan 26 & 491.54899 & 01:05:35 & 4$\times600$ & 240 & 2.6 & 27.968 & $7.1\pm3.8$   & $-14.08$ \\
     Jan 27 & 492.57173 & 01:38:15 & 4$\times600$ & 310 & 2.0 & 28.041 & $-3.5\pm3.0$  & $-14.07$  \\
     Jan 28 & 493.58339 & 01:54:59 & 4$\times600$ & 290 & 2.2 & 28.113 & $6.7\pm3.3$   & $-14.04$ \\

     Feb 03 & 499.60160 & 02:20:49 & 4$\times600$ & 270 & 2.3 & 28.543 & $-22.3\pm3.8$ & $-14.11$ \\
     Feb 05 & 501.58341 & 01:54:31 & 4$\times600$ & 250 & 2.6 & 28.684 & $-0.9\pm4.1$  & $-14.08$ \\
     Feb 06 & 502.58201 & 01:52:26 & 4$\times600$ & 310 & 2.1 & 28.756 & $4.3\pm3.1$   & $-14.08$ \\
     Feb 07 & 503.58042 & 01:50:06 & 4$\times600$ & 290 & 2.2 & 28.827 & $3.5\pm3.2$   & $-14.10$ \\
     Feb 10 & 506.59331 & 02:08:31 & 4$\times600$ & 290 & 2.1 & 29.042 & $4.2\pm3.2$   & $-14.03$ \\
     Feb 11 & 507.55697 & 01:16:08 & 4$\times600$ & 300 & 2.0 & 29.111 & $5.8\pm2.9$   & $-14.06$ \\
     Feb 12 & 508.58557 & 01:57:17 & 4$\times600$ & 310 & 2.2 & 29.185 & $16.1\pm3.3$  & $-14.04$ \\
     Feb 13 & 509.59071 & 02:04:38 & 4$\times600$ & 290 & 2.2 & 29.256 & $16.1\pm3.3$  & $-14.02$ \\
     Feb 14 & 510.58572 & 01:57:25 & 4$\times600$ & 230 & 2.9 & 29.328 & $10.9\pm4.3$  & $-14.03$ \\
     Feb 15 & 511.60258 & 02:21:39 & 4$\times600$ & 300 & 2.1 & 29.400 & $-9.4\pm3.2$  & $-13.98$ \\
     Feb 16 & 512.58834 & 02:01:06 & 4$\times600$ & 270 & 2.4 & 29.471 & $-16.5\pm3.5$ & $-14.04$ \\
     Feb 17 & 513.58759 & 01:59:59 & 4$\times600$ & 300 & 2.1 & 29.542 & $-16.9\pm3.1$ & $-14.02$ \\
\hline
\end{tabular}
\label{tab:log410}
\end{table*}

\begin{table*}
\caption[]{Same as Table~\ref{tab:log494} for CE~Boo=GJ~569A. 
Rotational cycles are now computed using ephemeris HJD$=2,454,100.0+14.7 E$. } 
\begin{tabular}{rcccccrrc}
\hline
Date & HJD          & UT      & $t_{\rm exp}$ & \sn\ & $\sigma_{\rm LSD}$ & Cycle & \Bl & \vr \\ 
     & (2,454,000+) & (h:m:s) &   (s)         &      &   (\ptt)           &       &  (G) & (\kms) \\
\hline
2008 Jan 19 & 484.66641 & 04:01:44 & 4$\times600$ & 130 & 6.2 & 26.168 & $-93.6\pm 14.5$ & $-7.31$ \\
     Jan 20 & 485.68258 & 04:24:54 & 4$\times600$ & 200 & 3.9 & 26.237 & $-92.2\pm 9.3$  & $-7.36$ \\
     Jan 23 & 488.65820 & 03:49:27 & 4$\times600$ & 230 & 3.3 & 26.439 & $-59.3\pm 7.8 $ & $-7.28$ \\
     Jan 24 & 489.65197 & 03:40:21 & 4$\times600$ & 190 & 3.7 & 26.507 & $-73.3\pm 8.8$  & $-7.29$ \\
     Jan 26 & 491.63667 & 03:18:05 & 4$\times600$ & 170 & 4.7 & 26.642 & $-110.2\pm 11.2$& $-7.31$ \\
     Jan 27 & 492.65802 & 03:48:42 & 4$\times600$ & 270 & 2.6 & 26.711 & $-90.5\pm 6.1$  & $-7.34$ \\
     Jan 28 & 493.67096 & 04:07:12 & 4$\times600$ & 280 & 2.5 & 26.780 & $-114.2\pm 6.1$ & $-7.31$ \\
     Jan 30 & 495.68078 & 04:21:06 & 4$\times600$ & 250 & 2.9 & 26.917 & $-111.0\pm 6.8$ & $-7.35$ \\

     Feb 03 & 499.69324 & 04:38:34 & 4$\times600$ & 260 & 2.7 & 27.190 & $-70.4\pm 6.5$  & $-7.42$ \\
     Feb 05 & 501.68094 & 04:20:37 & 4$\times600$ & 250 & 2.9 & 27.325 & $-53.8\pm 6.9$  & $-7.35$ \\
     Feb 07 & 503.67364 & 04:09:51 & 4$\times600$ & 260 & 2.7 & 27.461 & $-54.9\pm 6.4$  & $-7.35$ \\
     Feb 10 & 506.67974 & 04:18:16 & 4$\times600$ & 270 & 2.7 & 27.665 & $-91.0\pm 6.3$  & $-7.32$ \\
     Feb 11 & 507.68809 & 04:30:10 & 4$\times600$ & 280 & 2.4 & 27.734 & $-113.4\pm 5.9$ & $-7.36$ \\
     Feb 12 & 508.67116 & 04:05:40 & 4$\times600$ & 240 & 3.0 & 27.801 & $-115.0\pm 7.1$ & $-7.34$ \\
     Feb 13 & 509.67882 & 04:16:35 & 4$\times600$ & 230 & 3.1 & 27.869 & $-108.6\pm 7.4$ & $-7.33$ \\
     Feb 14 & 510.67193 & 04:06:32 & 4$\times600$ & 200 & 3.7 & 27.937 & $-106.7\pm 8.6$ & $-7.35$ \\
     Feb 15 & 511.68789 & 04:29:24 & 4$\times600$ & 270 & 2.7 & 28.006 & $-111.7\pm 6.4$ & $-7.31$ \\
     Feb 16 & 512.67287 & 04:07:39 & 4$\times600$ & 260 & 2.6 & 28.073 & $-85.5\pm 6.2$  & $-7.37$ \\
     Feb 17 & 513.67252 & 04:07:02 & 4$\times600$ & 270 & 2.6 & 28.141 & $-79.9\pm 6.2$  & $-7.34$ \\
\hline
\end{tabular}
\label{tab:log569}
\end{table*}

\begin{table*}
\caption[]{Same as Table~\ref{tab:log494} for OT~Ser=GJ~9520. 
Rotational cycles are now computed using ephemeris HJD$=2,454,100.0+3.40 E$. } 
\begin{tabular}{rcccccrrc}
\hline
Date & HJD          & UT      & $t_{\rm exp}$ & \sn\ & $\sigma_{\rm LSD}$ & Cycle & \Bl & \vr \\ 
     & (2,454,000+) & (h:m:s) &   (s)         &      &   (\ptt)           &       &  (G) & (\kms) \\
\hline
2007 Jul 26 & 308.42253 & 22:05:53 & 4$\times900$ & 270 & 2.7 & 61.301 & $80.0\pm5.7$ & $6.89$ \\
     Jul 28 & 310.38549 & 21:12:45 & 4$\times900$ & 340 & 2.0 & 61.878 & $55.0\pm4.6$ & $6.75$ \\
     Jul 29 & 311.43298 & 22:21:15 & 4$\times900$ & 370 & 1.8 & 62.186 & $73.4\pm4.1$ & $6.87$ \\
     Jul 30 & 312.41904 & 22:01:17 & 4$\times900$ & 380 & 1.8 & 62.476 & $74.9\pm3.8$ & $6.82$ \\
     Jul 31 & 313.41995 & 22:02:42 & 4$\times900$ & 230 & 3.2 & 62.771 & $67.0\pm7.0$ & $6.82$ \\
     Aug 02 & 315.42156 & 22:05:14 & 4$\times900$ & 350 & 1.9 & 63.359 & $80.9\pm4.2$ & $6.90$ \\
     Aug 03 & 316.42155 & 22:05:20 & 4$\times900$ & 350 & 1.9 & 63.653 & $67.8\pm4.2$ & $6.84$ \\
     Aug 04 & 317.42236 & 22:06:36 & 4$\times900$ & 230 & 3.1 & 63.948 & $55.5\pm7.1$ & $6.77$ \\
     Aug 09 & 322.41667 & 21:58:58 & 4$\times900$ & 330 & 2.1 & 65.417 & $77.7\pm4.5$ & $6.87$ \\
     Aug 10 & 323.41738 & 22:00:06 & 4$\times900$ & 250 & 2.8 & 65.711 & $70.5\pm6.1$ & $6.82$ \\
     Aug 14 & 327.35676 & 20:33:15 & 4$\times900$ & 270 & 2.7 & 66.870 & $54.5\pm6.0$ & $6.79$ \\
\hline
2008 Jan 19 & 484.74579 & 05:56:56 & 4$\times600$ & 150 & 5.2 & 113.160 & $52.7\pm10.6$ & $6.78$ \\
     Jan 20 & 485.71587 & 05:13:45 & 4$\times600$ & 190 & 3.9 & 113.446 & $1.4\pm8.1$   & $6.80$ \\
     Jan 23 & 488.69155 & 04:38:25 & 4$\times600$ & 180 & 4.1 & 114.321 & $8.9\pm8.7$   & $6.81$ \\
     Jan 24 & 489.68897 & 04:34:36 & 4$\times600$ & 160 & 4.5 & 114.614 & $69.1\pm9.6$  & $6.76$ \\
     Jan 26 & 491.67101 & 04:08:31 & 4$\times600$ & 200 & 3.6 & 115.197 & $47.5\pm7.6$  & $6.78$ \\
     Jan 27 & 492.69173 & 04:38:14 & 4$\times600$ & 260 & 2.8 & 115.498 & $22.4\pm5.8$  & $6.79$ \\
     Jan 28 & 493.70473 & 04:56:51 & 4$\times600$ & 280 & 2.4 & 115.796 & $90.9\pm5.3$  & $6.81$ \\
     Jan 30 & 495.71492 & 05:11:18 & 4$\times600$ & 250 & 2.8 & 116.387 & $-6.6\pm5.9$  & $6.80$ \\

     Feb 03 & 499.72743 & 05:28:51 & 4$\times600$ & 240 & 2.9 & 117.567 & $47.7\pm6.2$  & $6.72$ \\
     Feb 05 & 501.71560 & 05:11:36 & 4$\times600$ & 260 & 2.7 & 118.152 & $86.3\pm5.7$  & $6.71$ \\
     Feb 07 & 503.70655 & 04:58:20 & 4$\times600$ & 280 & 2.5 & 118.737 & $79.7\pm5.4$  & $6.75$ \\
     Feb 10 & 506.71299 & 05:07:16 & 4$\times600$ & 260 & 2.7 & 119.621 & $52.6\pm5.6$  & $6.77$ \\
     Feb 11 & 507.72197 & 05:20:05 & 4$\times600$ & 280 & 2.5 & 119.918 & $108.3\pm5.4$ & $6.71$ \\
     Feb 12 & 508.70420 & 04:54:23 & 4$\times600$ & 240 & 3.0 & 120.207 & $56.9\pm6.2$  & $6.75$ \\
     Feb 13 & 509.62257 & 02:56:43 & 4$\times600$ & 230 & 3.1 & 120.477 & $9.8\pm6.6$   & $6.78$ \\
     Feb 13 & 509.75390 & 06:05:49 & 4$\times600$ & 270 & 2.6 & 120.516 & $27.8\pm5.5$  & $6.77$ \\
     Feb 14 & 510.70531 & 04:55:45 & 4$\times600$ & 220 & 3.3 & 120.796 & $92.5\pm6.8$  & $6.78$ \\
     Feb 15 & 511.72139 & 05:18:47 & 4$\times600$ & 270 & 2.5 & 121.094 & $104.7\pm5.4$ & $6.73$ \\
     Feb 16 & 512.70638 & 04:57:03 & 4$\times600$ & 280 & 2.5 & 121.384 & $10.3\pm5.4$  & $6.74$ \\
     Feb 17 & 513.70577 & 04:56:03 & 4$\times600$ & 270 & 2.5 & 121.678 & $76.0\pm5.4$  & $6.77$ \\
\hline
\end{tabular}
\label{tab:log9520}
\end{table*}

\begin{table*}
\caption[]{Same as Table~\ref{tab:log494} for GJ~182. 
Rotational cycles are now computed using ephemeris HJD$=2,454,100.0+4.35 E$. } 
\begin{tabular}{rcccccrrc}
\hline
Date & HJD          & UT      & $t_{\rm exp}$ & \sn\ & $\sigma_{\rm LSD}$ & Cycle & \Bl  & \vr \\ 
     & (2,454,000+) & (h:m:s) &   (s)         &      &   (\ptt)           &       &  (G) & (\kms)\\
\hline
2007 Jan 21 & 122.37018 & 20:48:02 & 4$\times900$ & 210 & 3.2 & 5.143 & $-84.4\pm8.0$  & $19.23$ \\
     Jan 26 & 127.28464 & 18:45:22 & 4$\times900$ & 190 & 3.6 & 6.272 & $-40.6\pm9.2$  & $19.22$ \\
     Jan 27 & 128.28237 & 18:42:12 & 4$\times900$ & 240 & 2.6 & 6.502 & $-37.9\pm6.4$  & $19.59$ \\
     Jan 30 & 131.27339 & 18:29:36 & 2$\times900$ & 55 & 12.7 & 7.189 & $-65.3\pm36.4$ & $19.22$ \\
     Feb 01 & 133.29669 & 19:03:23 & 4$\times900$ & 250 & 2.4 & 7.654 & $18.6\pm5.9$   & $19.66$ \\
     Feb 03 & 135.28638 & 18:48:47 & 4$\times900$ & 190 & 3.4 & 8.112 & $-84.2\pm8.7$  & $19.21$ \\
     Feb 08 & 140.29936 & 19:08:04 & 4$\times900$ & 250 & 2.5 & 9.264 & $-42.6\pm6.5$  & $19.24$ \\
\hline
\end{tabular}
\label{tab:log182}
\end{table*}

\begin{table*}
\caption[]{Same as Table~\ref{tab:log494} for GJ~49. 
Rotational cycles are now computed using ephemeris HJD$=2,454,300.0+18.6 E$. } 
\begin{tabular}{rcccccrrc}
\hline
Date & HJD          & UT      & $t_{\rm exp}$ & \sn\ & $\sigma_{\rm LSD}$ & Cycle & \Bl  & \vr \\ 
     & (2,454,000+) & (h:m:s) &   (s)         &      &   (\ptt)           &       &  (G) & (\kms) \\
\hline
2007 Jul 27 & 308.61963 & 02:49:20 & 4$\times900$ & 330 & 2.1 & 0.463 & $-4.6\pm4.1$  & $-6.00$ \\
     Jul 28 & 309.63521 & 03:11:41 & 4$\times900$ & 450 & 1.5 & 0.518 & $-12.0\pm2.8$ & $-6.02$ \\
     Jul 29 & 310.63481 & 03:10:59 & 4$\times900$ & 410 & 1.6 & 0.572 & $-20.9\pm3.1$ & $-6.02$ \\
     Jul 30 & 311.63656 & 03:13:25 & 4$\times700$ & 370 & 1.8 & 0.626 & $-28.4\pm3.5$ & $-6.01$ \\
     Jul 31 & 312.63031 & 03:04:19 & 4$\times700$ & 380 & 1.7 & 0.679 & $-29.2\pm3.3$ & $-6.06$ \\
     Aug 01 & 313.63144 & 03:05:51 & 4$\times700$ & 300 & 2.2 & 0.733 & $-21.7\pm4.3$ & $-6.01$ \\

     Aug 03 & 315.63927 & 03:16:55 & 4$\times700$ & 320 & 2.5 & 0.841 & $-5.9\pm5.6$  & $-6.02$ \\
     Aug 04 & 316.63804 & 03:15:03 & 4$\times700$ & 380 & 1.7 & 0.894 & $-1.6\pm3.4$  & $-6.04$ \\
     Aug 05 & 317.60746 & 02:30:56 & 4$\times700$ & 350 & 1.9 & 0.947 & $2.9\pm3.7$   & $-6.01$ \\
     Aug 09 & 321.61210 & 02:37:15 & 4$\times700$ & 340 & 2.0 & 1.162 & $-10.2\pm3.7$ & $-6.02$ \\
     Aug 10 & 322.63155 & 03:05:10 & 4$\times700$ & 330 & 2.1 & 1.217 & $-10.6\pm3.9$ & $-6.02$ \\
     Aug 11 & 323.63437 & 03:09:08 & 4$\times700$ & 280 & 2.5 & 1.271 & $-16.4\pm4.7$ & $-6.04$ \\
     Aug 15 & 327.62517 & 02:55:34 & 4$\times700$ & 350 & 2.0 & 1.485 & $-10.3\pm3.8$ & $-6.03$ \\
     Aug 18 & 330.61820 & 02:45:17 & 4$\times700$ & 370 & 1.8 & 1.646 & $-28.1\pm3.5$ & $-6.03$ \\
     Aug 19 & 331.57357 & 01:40:56 & 4$\times700$ & 340 & 1.9 & 1.698 & $-20.2\pm3.6$ & $-6.06$ \\
     Aug 31 & 343.56027 & 01:20:59 & 4$\times700$ & 280 & 2.4 & 2.342 & $-5.9\pm4.7$  & $-6.00$ \\
\hline
\end{tabular}
\label{tab:log49}
\end{table*}

Least-squares deconvolution \citep[LSD, ][]{Donati97} was applied to all spectra to 
extract the polarisation signal from most photospheric atomic lines and compute a 
mean Zeeman signature corresponding to an average photospheric profile (centred at 
700~nm and with an effective Land\'e factor of 1.2).  The line list used in this 
process is derived from an Atlas9 local thermodynamic equilibrium model \citep{Kurucz93} 
matching the properties of our sample, and includes about 5,000 moderate 
to strong atomic lines (i.e., with a relative depth larger than 40\% prior to any 
macroscopic broadening).  The resulting multiplex gain in \sn\ is about 15 
(see Table~\ref{tab:log494}).  Zeeman signatures are detected in most cases on all 
stars.  Radial velocities are obtained from Gaussian fits to all LSD unpolarised 
profiles.  Average unpolarised LSD profiles of all stars are shown in Fig.~\ref{fig:lsd}.  

\begin{figure}
\includegraphics[scale=0.37,angle=-90]{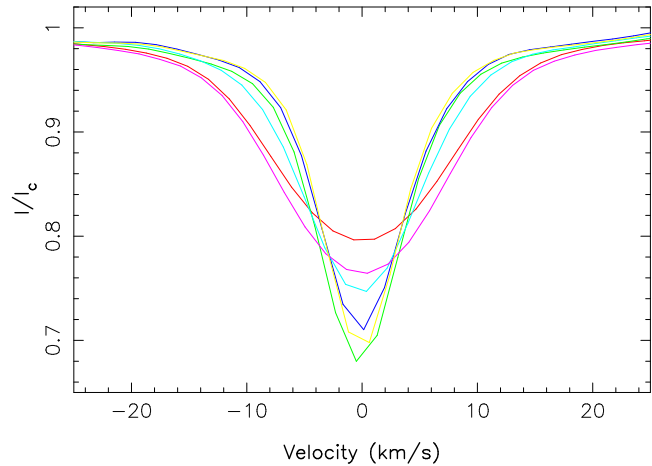}
\caption[]{Average unpolarised LSD profiles for DT~Vir (red line), DS~Leo (green), CE~Boo (dark blue), 
OT~Ser (light blue), GJ~182 (pink) and GJ~49 (yellow).  All profiles are centred on the rest velocity.  }  
\label{fig:lsd}
\end{figure}

\section{Magnetic modelling}
\label{sec:mod}

The magnetic model we use to describe the time series of rotationally modulated LSD 
Stokes $V$ profiles is described in M08.  We recall it briefly here and refer the reader 
to M08 for further details.  

To describe the magnetic field, we use the same description as \citet{Donati06b}.  The 
field is decomposed into its poloidal and toroidal components, both expressed as 
spherical harmonics expansions.  The imaging process is based on the principles of maximum 
entropy image reconstruction, with entropy (i.e., quantifying the amount of reconstructed 
information) being calculated from the coefficients of the spherical harmonics expansions.  
Starting from a null magnetic field, we iteratively improve our magnetic model by comparing 
the synthetic Stokes $V$ profiles with the observed ones, until we reach an optimal field 
topology that reproduces the data at a given \chisq\ level (i.e., usually down to noise 
level, corresponding to a unit reduced \chisq\ level $\chinu=1$).  The inversion problem 
being partly ill-posed, we use the entropy function to select the magnetic field with 
lowest information content among all those reproducing the data equally well.  
Given that most stars considered here rotate no more than moderately, we limit 
spherical harmonics expansions to $\ell<10$, usually up to 8 for moderate rotators 
(e.g., DT~Vir) and up to 5 only for the slower ones.  In all cases (even the slowest 
rotators), we need to set $\ell>3$ to reproduce successfully the data at noise level.  

To compute the synthetic profiles corresponding to a given magnetic topology, we divide 
the surface of the star into a grid of elementary surface cells (typically 5,000), in 
which the 3 components of the magnetic field (in spherical coordinates) are estimated 
directly from the spherical harmonics expansions used to describe the field.  Using 
Unno-Rachkovsky's equations \citep[e.g.,][]{Landi92}, we compute the contribution of each 
grid cell to the 
Stokes $V$ profiles and integrate all contributions from the visible stellar hemisphere 
at each observed rotation phase.  The free parameters in Unno-Rachkovsky's equations 
(describing the shape of the unpolarised line profile from a non-magnetic grid cell) 
are obtained by fitting the Stokes $I$ LSD profiles of a very-slowly rotating and 
weakly active star of similar spectral type (e.g., GJ~205).  

Reproducing both the amplitude and shape of LSD Stokes $V$ profiles in the particular 
case of stars with strong fields and sharp lines requires that we introduce a 
filling factor (called $f$) describing the fractional amount of flux producing circular 
polarisation (assumed constant over the whole star).  At first glance, this may 
seem in contradiction with the fact that we are mostly sensitive to large-scale fields, 
i.e., fields whose spatial coherency is much larger than the size of our grid cells;  
however, large-scale fields can potentially be also structured on a small-scale, e.g., 
with convection compressing the field into a small section of each cell but keeping the 
flux constant over the cell surface.  We suspect that this is the case here.  
In practice, it means that the circularly polarised flux that we get from each grid cell 
is given by $f V_{\rm loc}$ where $V_{\rm loc}$ is the Stokes $V$ profile derived from 
Unno-Rachkovsky's equations for a magnetic strength of $B/f$.  In practice, this 
simple model ensures that both the width and amplitude of Stokes $V$ signatures can 
be fitted simultaneously;  typical values of $f$ range from 0.10 to 0.15 in active 
mid-M dwarfs with sharp lines (M08).  For stars rotating more rapidly and/or 
hosting intrinsically weak fields, different values of $f$ produce very similar fits 
and undistinguishable magnetic flux maps, in which case we arbitrarily set $f=1$.  
This model has proved rather successful at reproducing the observed times series of 
Stokes $V$ profiles in mid-M dwarfs (M08) and classical T~Tauri stars \citep{Donati08};  
we therefore use it again for the present study.  

We can also implement differential rotation for computing the 
synthetic Stokes $V$ profiles corresponding to our magnetic model.  For this 
purpose, we use a Sun-like surface rotation pattern with the rotation rate varying 
with latitude $\theta$ as $\omeq-\dom\sin^2 \theta$,  \omeq\ being the angular 
rotation rate at the equator and \dom\ the difference in angular rotation rate 
between the equator and the pole.  By carrying out reconstructions (at constant 
information content) for a range of \omeq\ and \dom\ values, we can investigate 
how the fit quality varies with differential rotation;  differential rotation is 
detected when the \chisq\ of the fit to the data shows a well defined minimum in 
the explored \omeq--\dom\ domain, with the position of the minimum and the 
curvature of the \chisq\ surface at this point yielding the optimal \omeq\ and 
\dom\ values and respective error bars \citep{Donati03b}.

\section{DT Vir = GJ 494A = HIP 63942}
\label{sec:gl494}

DT~Vir is a magnetically active M0.5 dwarf showing significant 
rotational broadening in spectral lines \citep[$\vsini\simeq10$~\kms,][]{Beuzit04};  
photometric variability suggests a short rotation period \prot\ of about 2.9~d 
in good agreement with its membership to the young galactic 
disc \citep{Kiraga07}.  The Hipparcos distance is $11.43\pm0.20$~pc.  It belongs to 
a distant binary system with an astrometric period of about 14.5~yr \citep{Heintz94} 
and a semi-major axis of 5--6~AU;  the companion is about 4.4 magnitudes fainter in K 
and is either a very-low-mass star or a young brown dwarf \citep[][depending on the 
exact age of the system]{Beuzit04}.  Using the mass-luminosity relations of 
\citet{Delfosse00}, we estimate that the mass of DT~Vir is $0.59\pm0.02$~\msun.  
Given \vsini\ and \prot, we infer that $\rstar\sin i$ is about 0.6~\rsun, i.e., already 
10\% larger than the radius expected from theoretical models (see Table~\ref{tab:sample});  
we therefore set $i=60\degr$ in the imaging process (the result being weakly sensitive 
to variations of $i$ of $\pm10\degr$).  

Stokes $V$ data were collected at 2 epochs, providing only moderate coverage of the 
rotation cycle at the first epoch but a dense and redundant coverage at the second 
epoch (see Table~\ref{tab:log494}).  Stokes $V$ signatures are clearly detected at 
all time, eventhough the corresponding longitudinal fields are usually low (ranging 
from --30~G to 70~G), much lower than those reported on 
mid-M dwarfs in particular (M08).  
The projected rotation velocity that we derive from the Stokes $I$ profiles is 
$\vsini=11\pm1$~\kms, in good agreement with the estimate of \citet{Beuzit04}.  
The RV we measure are different at both epochs, equal to 
$-13.25$~\kms\ and $-13.68$~\kms\ respectively (with an absolute accuracy of about 
0.10~\kms);  this difference likely reflects the binary motion.  
Moreover, the relative dispersion about the mean RV, respectively 
equal to 0.14~\kms\ and 0.07~\kms, is larger than the internal RV accuracy of NARVAL 
\citep[about 0.03~\kms, e.g.,][]{Moutou07} and likely reflects the intrinsic activity 
RV jitter of DT~Vir;  in 2007 (i.e., when the internal RV dispersion is largest), we 
find that RVs correlate reasonably well with longitudinal fields, suggesting that the RV 
fluctuations are indeed due to the magnetic activity.  Given the moderate strength of 
the field and the significant rotational broadening of DT~Vir, there is no need of 
adjusting $f$ to optimise the fit quality.  

%% fig0:
%% \begin{figure}
%% \includegraphics[scale=0.33,angle=-90]{fig/mdw_494beff.ps}
%% \caption[]{Longitudinal fields (with $\pm$1$\sigma$ error bars) of DT~Vir 
%% as a function of rotational cycle, for both the 2007 (green) and 2008 (red) runs.  }
%% \label{fig:494beff}
%% \end{figure}

Assuming solid body rotation, we find that the rotation period providing the best 
fit to the data is close to that derived from photometric variations but slightly 
different for each of the 2 data sets, about 2.90~d for the 2008 data and 2.80~d 
for the 2007 data;  we therefore selected $\prot=2.85$~d as the mean rotation period 
with which we phased all data.  Using this value of \prot\ yields a slightly chaotic 
phase dependence for the 2008 \Bl\ data, with points at nearby phases but different 
cycles (e.g., on Dec~28 and Feb~06, i.e., at rotation cycles 127.269 and 141.276) 
showing discrepant field values;  we suspect it 
indicates significant surface differential rotation on DT~Vir.  
Further confirmation comes from our finding that the 2008 data 
cannot be fitted down to $\chinu=1$ for solid body rotation;  using 
differential rotation, we are able to fit to the data down to noise level, with 
the \chinu\ surface (at given information content) showing a clear minimum.  
The differential rotation parameters we obtain are $\omeq=2.200\pm0.003$~\rpd\ 
and $\dom=0.060\pm0.006$~\rpd, corresponding to rotation periods at the equator 
and pole of 2.85~d and 2.94~d respectively (bracketing the estimate of 
\citealt{Kiraga07}).  The photospheric shear of DT~Vir is very similar to 
that of the Sun, with the equator lapping the pole by one rotation cycle every 
$105\pm10$~d.  

The optimal maximum entropy fit to the Stokes $V$ data that we obtain (including 
the effect of differential rotation) is shown in Fig.~\ref{fig:494fit} and 
corresponds to $\chinu=1$, i.e., to a \chinu\ improvement over a non-magnetic 
model of $\times31$ and $\times22$ for each epoch respectively.  The reconstructed 
magnetic maps are shown in Fig.~\ref{fig:494map}.   

%% fig1:
\begin{figure*}
\center{\includegraphics[scale=0.55,angle=-90]{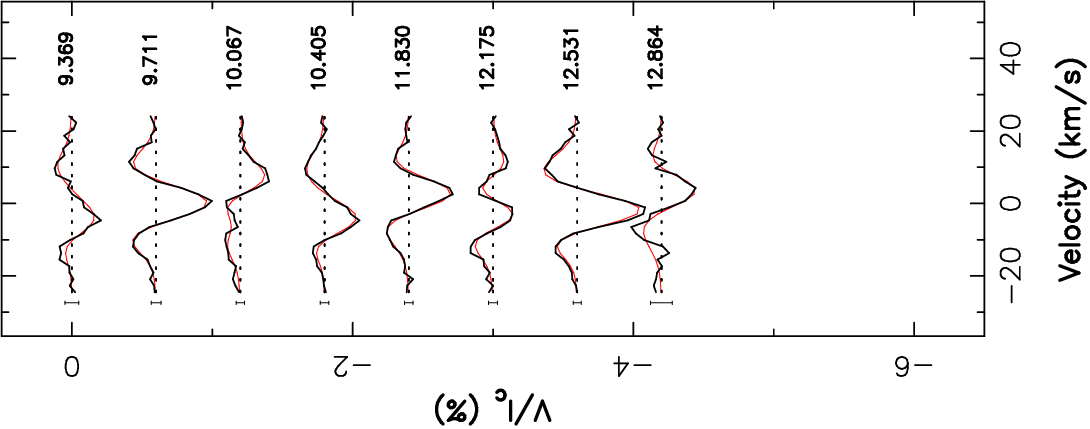}\hspace{1mm}
        \includegraphics[scale=0.55,angle=-90]{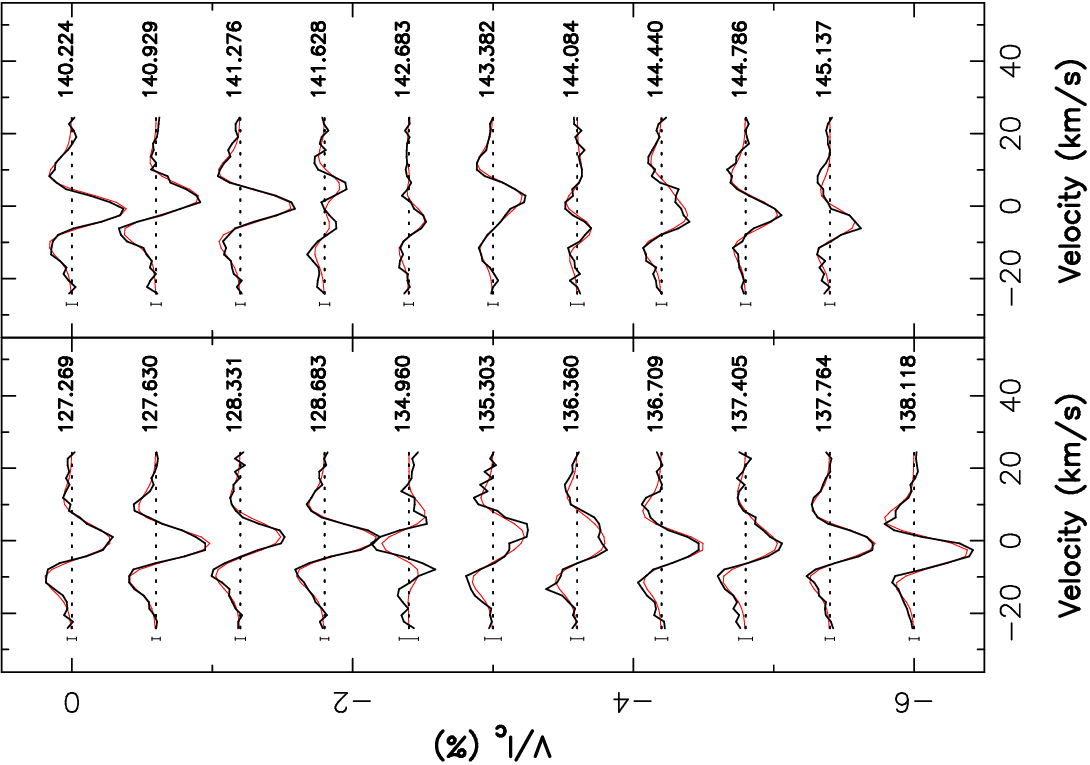}}
\caption[]{Stokes $V$ LSD profiles of DT~Vir=GJ~494A (thick black line) along with 
the maximum entropy fit (thin red line) to the data, for both the 2007 (left) and 
2008 (right) runs.  The rotational cycle of each observation, along with 3$\sigma$ 
error bars, are shown next to each profile.  }
\label{fig:494fit}
\end{figure*}

%% fig2:
\begin{figure*}
\center{\includegraphics[scale=0.65]{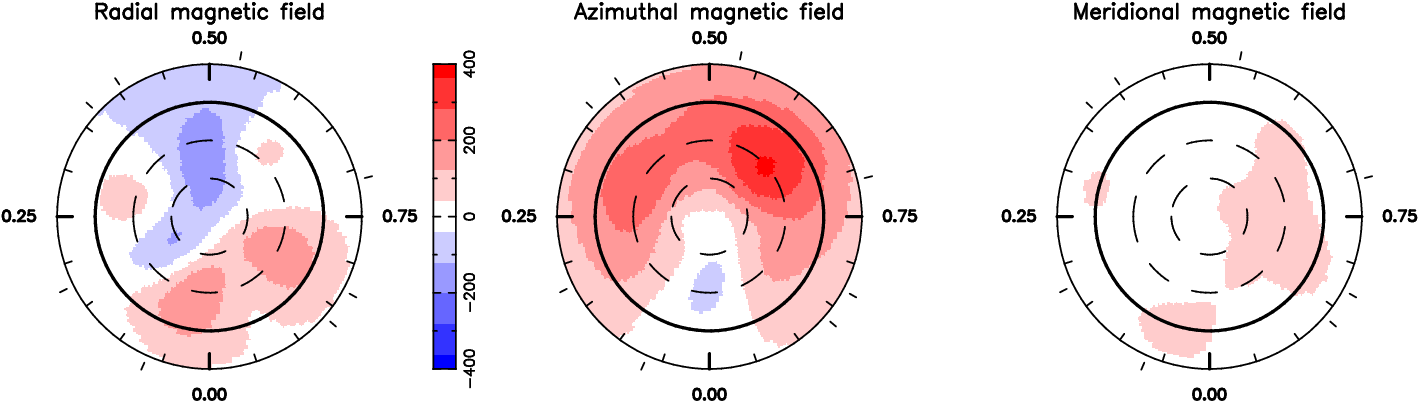}}
\center{\includegraphics[scale=0.65]{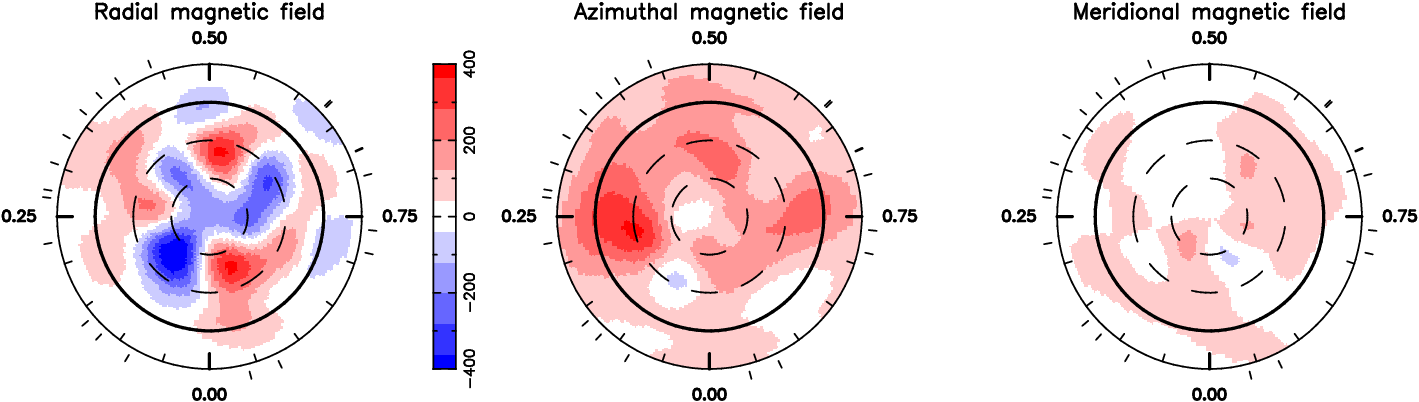}}
\caption[]{Magnetic topologies of DT~Vir=GJ~494A in 2007 (top) and 2008 (bottom),  
reconstructed from a fit to the complete series of LSD Stokes $V$ profiles.  
The three components of the field in spherical coordinates are displayed (from 
left to right), with magnetic fluxes labelled in G.  The star is shown in flattened 
polar projection down to latitudes of $-30\degr$, with the equator depicted as a 
bold circle and parallels as dashed circles.   Radial ticks around each plot 
indicate the phases of observation. }
\label{fig:494map}
\end{figure*}

As obvious from Fig.~\ref{fig:494map}, the magnetic topologies we derive contain 
a significant amount of toroidal field (62\% and 47\% of the reconstructed 
magnetic energy in 2007 and 2008 respectively);  at both epochs, the 
reconstructed toroidal field shows up as a ring of counterclockwise field 
encircling the whole star.  The poloidal field is more complex, especially in 
2008 when more than 50\% of the reconstructed poloidal field energy concentrates 
in orders with $\ell>3$ (with dipole modes containing only 10\%).  
The intrinsic evolution of the large-scale field topology between 2007 and 2008 
is straightforwardly visible in the image, especially on the poloidal field 
component whose spatial structure was much simpler in 2007 (64\% of the 
poloidal field energy in dipole modes).  At both epochs, the reconstructed 
poloidal field is mostly non axisymmetric (less than 20\% of the energy 
concentrating in $m<\ell/2$ modes).  This information is summarised in 
Table~\ref{tab:syn}.  

%% fig0:
%% \begin{figure}
%% \includegraphics[scale=0.33,angle=-90]{fig/mdw_410beff.ps}
%% \caption[]{Same as Fig.~\ref{fig:494beff} for DS~Leo}.  
%% \label{fig:410beff}
%% \end{figure}

\section{DS Leo = GJ 410 = HD 95650 = HIP 53985}
\label{sec:gl410}

DS~Leo is a single M0 dwarf with sharp spectral lines, located at an Hipparcos 
distance of $11.66\pm0.18$~pc from the Sun.  Its RV is equal to 
$-13.90\pm0.10$~\kms\ \citep{Nidever02}.  
Photometric variability was studied by \citet{Fekel00} who 
detected cyclic variability at periods of 13.99~d and 15.71~d on different 
observing seasons and interpreted it as caused by stellar rotation (about 14~d) 
coupled to surface differential rotation (modulating the observed photometric 
period as spots migrate to different latitudes).  Using the mass-luminosity relations of
\citet{Delfosse00}, we estimate that the mass of DS~Leo is $0.58\pm0.02$~\msun, i.e.\ 
very similar to that of DT~Vir.  The large amplitude rotational modulation that 
we observe for Stokes $V$ profiles suggest that the inclination angle is not small;  
we therefore set $i=60\degr$ in the following imaging process.  

Stokes $V$ data were collected at 2 epochs (same runs as for DT~Vir, see 
Table~\ref{tab:log410}), providing again 
partial coverage of the rotation cycle at the first epoch but a dense and redundant 
coverage at the second epoch.  Stokes $V$ signatures are detected in almost all 
spectra, with longitudinal fields never exceeding strengths of 35~G.  
The RV we measure ($-14.05\pm0.10$~\kms\ at both epochs, with an internal dispersion 
of 0.03~\kms) is in reasonbly good agreement with that of \citet{Nidever02}. 
The rotational broadening of DS~Leo is small, only slightly larger than that of 
GJ~205 \citep[having $\vsini=1.0-1.5$~\kms, ][]{Reiners07b};  using 
$\vsini=2\pm1$~\kms\ provides a good fit to the Stokes $I$ profiles and is 
compatible with the radius expected from theoretical models (0.52~\rsun, see 
Table~\ref{tab:sample}), the rotation period of \citet{Fekel00} and the inclination 
angle we assumed ($i=60\degr$).  
As for DT~Vir, we do not need to use $f$ for modelling the profiles of DS~Leo.  

The solid-body-rotation period providing the best fit to the data is close to 14~d 
at both epochs;  we therefore used it to phase all spectra.  As for DT~Vir, the 
\Bl\ curve in the 2008 data is showing apparently discrepant points for spectra 
collected at nearby phases but different cycles (e.g., Jan~03 and Feb~14, at rotation 
cycles 26.336 and 29.328) as a likely result of the presence of surface differential 
rotation.  This is confirmed by the fact that the full Stokes $V$ 2008 data set 
cannot be fitted down to $\chinu=1$ when assuming solid body rotation.  
Proceeding as for DT~Vir, we obtain that $\omeq=0.465\pm0.004$~\rpd\ 
and $\dom=0.076\pm0.020$~\rpd\ at the surface of DS~Leo, i.e., that the rotation 
periods at the equator and the pole are respectively equal to 13.5~d and 16.1~d 
(bracketing both photometric periods of \citealt{Fekel00}).  The photospheric shear 
of DS~Leo is thus very similar to that of DT~Vir, with the equator lapping the pole by 
one cycle every $83_{-20}^{+30}$~d.  

The optimal maximum entropy fit to the Stokes $V$ data that we obtain (including 
the effect of differential rotation) is shown in Fig.~\ref{fig:410fit} and 
corresponds to $\chinu=0.9$, i.e., to a \chinu\ improvement over a non-magnetic 
model of $\times15$ and $\times8$ for each epoch respectively.  The reconstructed 
magnetic maps are shown in Fig.~\ref{fig:410map}.   

%% fig3:
\begin{figure*}
\center{\includegraphics[scale=0.55,angle=-90]{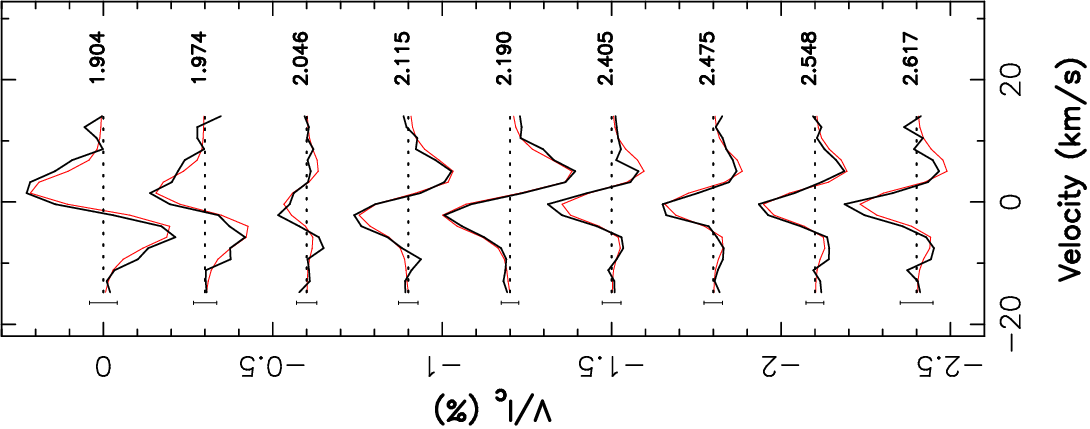}\hspace{1mm}
        \includegraphics[scale=0.55,angle=-90]{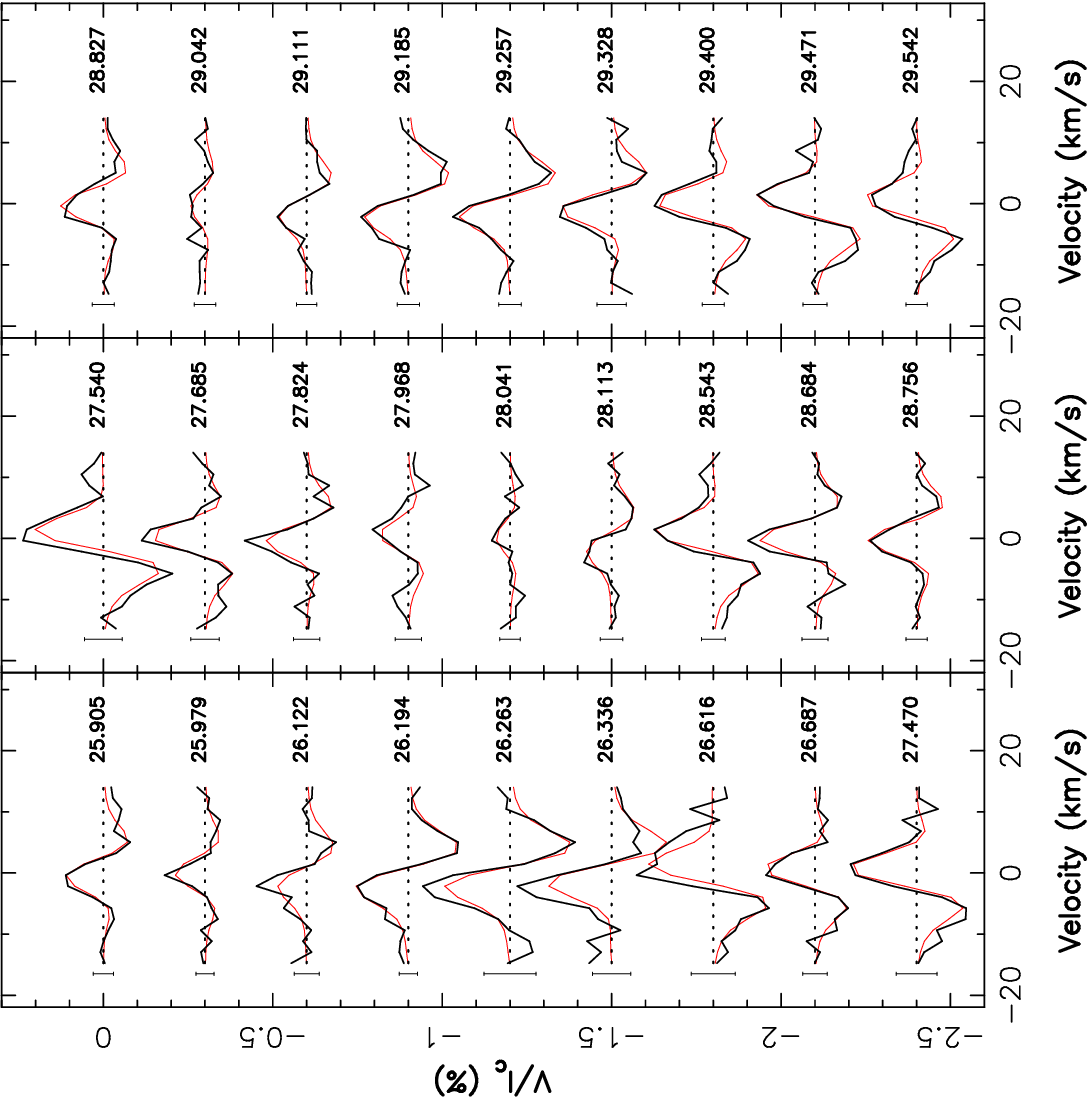}}
\caption[]{Same as Fig.~\ref{fig:494fit} for DS~Leo=GJ~410.  } 
\label{fig:410fit}
\end{figure*}

%% fig4:
\begin{figure*}
\center{\includegraphics[scale=0.65]{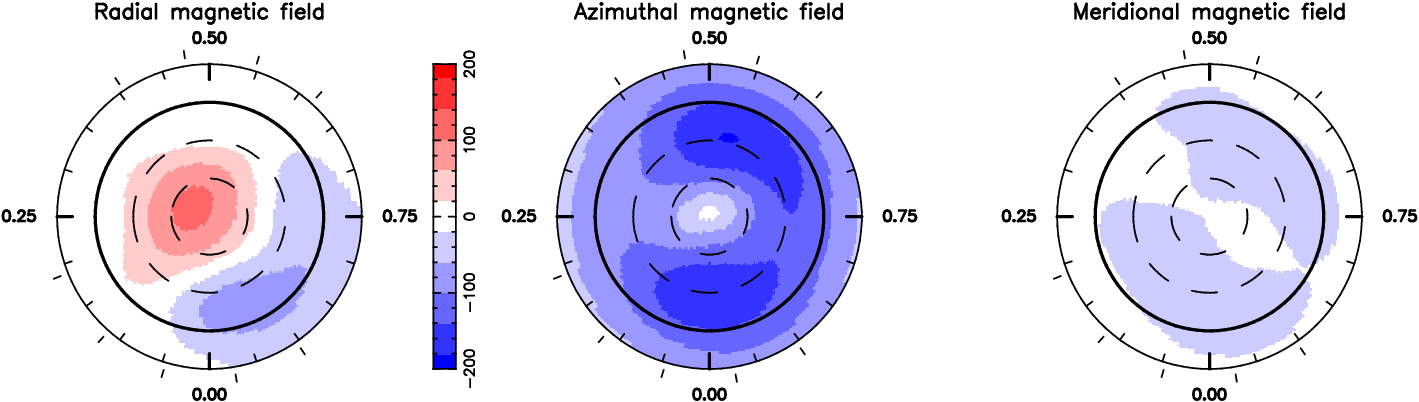}}
\center{\includegraphics[scale=0.65]{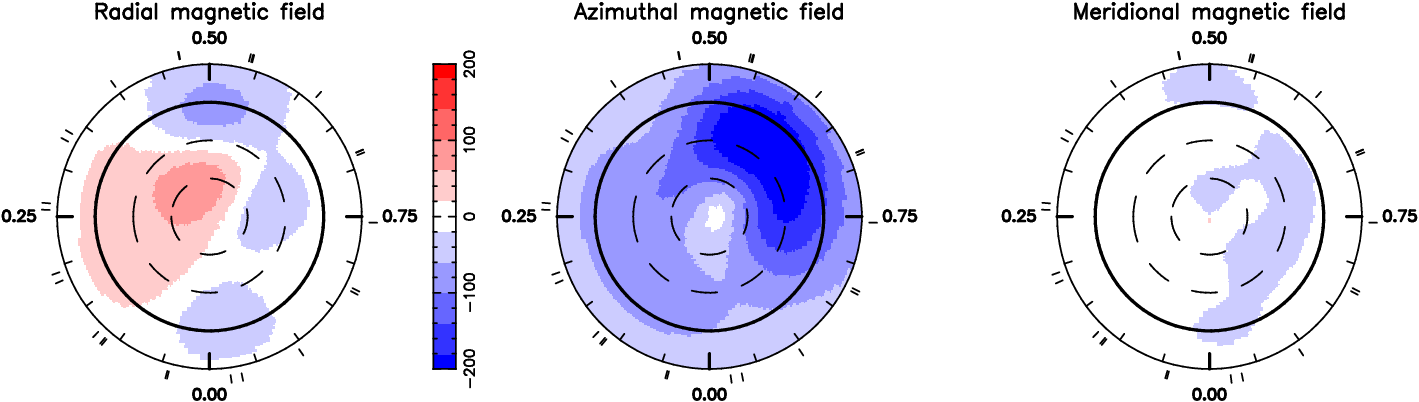}}
\caption[]{Same as Fig.~\ref{fig:494map} for DS~Leo=GJ~410.  }
\label{fig:410map}
\end{figure*}

%% fig3:
\begin{figure}
\center{\includegraphics[scale=0.55,angle=-90]{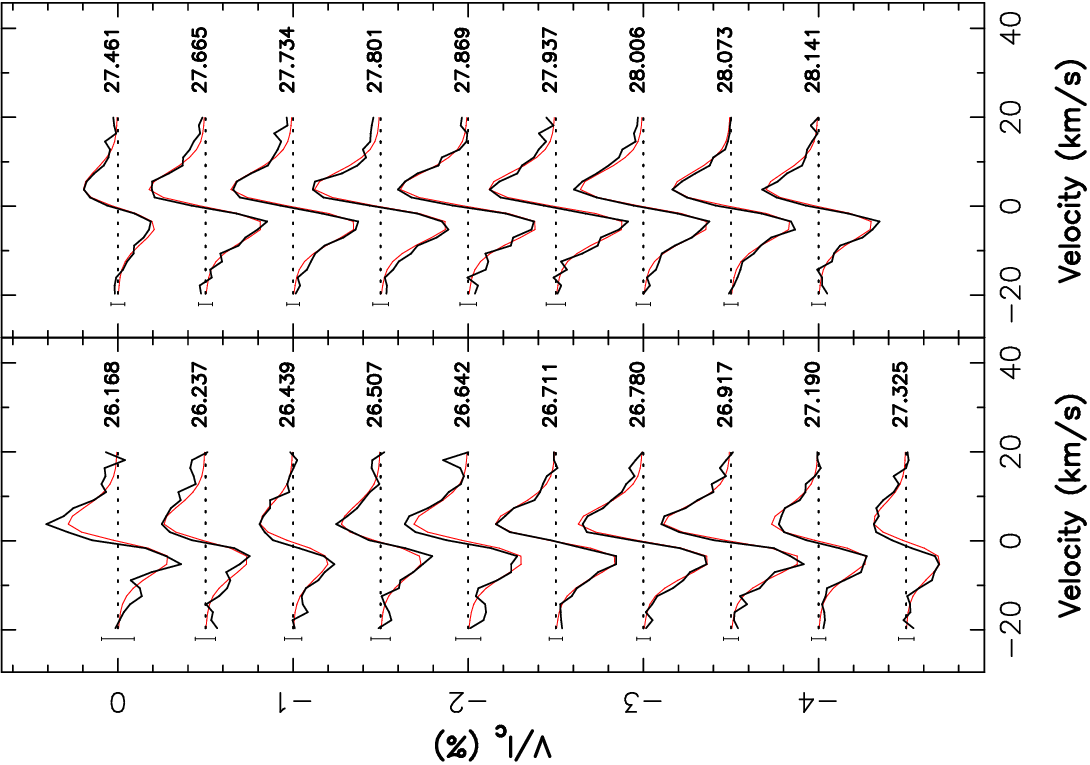}}
\caption[]{Same as Fig.~\ref{fig:494fit} for CE~Boo=GJ~569A.  }
\label{fig:569fit}
\end{figure}

%% fig4:
\begin{figure*}
\center{\includegraphics[scale=0.65]{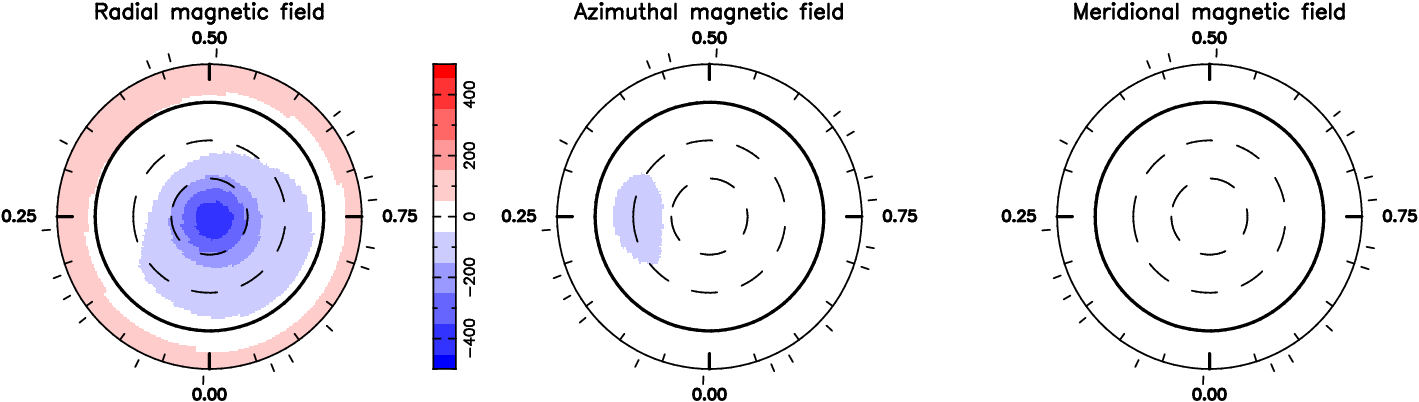}}
\caption[]{Same as Fig.~\ref{fig:494map} for CE~Boo=GJ~569A.  }
\label{fig:569map}
\end{figure*}

The magnetic topologies we derive are predominantly toroidal (more than 80\% of the 
reconstructed magnetic energy at both epochs), with the reconstructed toroidal field 
showing up as a ring of clockwise field encircling the star.  
The poloidal field is much simpler than that of DT~Vir (partly because of the lower 
spatial resolution resulting from the smaller \vsini) and consists mostly in a 
dipole (containing more than 50\% of the poloidal field energy) evolving from a 
mainly axisymmetric to a mainly non-axisymmetric configuration between 2007 and 2008.  
Eventhough the fractional energy stored in the various field components remained 
grossly stable (see Table~\ref{tab:syn}), the magnetic topology underwent significant 
temporal evolution between both epochs, e.g., with the toroidal field ring showing 
2 extrema across the star in 2007 while it only shows one in 2008.

\section{CE Boo = GJ 569A = HIP 72944}
\label{sec:gj569}

CE~Boo is a young and active M2.5 dwarf with sharp spectral lines, located at an Hipparcos 
distance of $9.81\pm0.16$~pc.  
It is the brightest member of a multiple (possibly quadruple) system with an estimated 
age of only $\simeq100$~Myr \citep{Simon06}.  The companions are located about 
5\arcsec\ (i.e., 50~AU) away and consist of at least a brown dwarf binary and possibly 
even a triple.  Using the mass-luminosity relations of \citet{Delfosse00}, we estimate a 
mass of $0.48\pm0.02$~\msun.  
The current RV is equal to $-7.21\pm0.10$ \citep{Nidever02}.  
From photometric variability, \citet{Kiraga07} find that the rotation period is 13.7~d;  
this is surprisingly long for a star as young as CE~Boo, even longer than the average 
period for M dwarfs of the young galactic disc (whose age is typically a few Gyr, 
\citealt{Kiraga07}).  The activity of CE~Boo is slightly larger than that of DS~Leo (but 
smaller than that of DT~Vir, see Table~\ref{tab:sample}), in agreement with what is 
expected for a star with a similar period and a later spectral type.  The sharp lines 
of CE~Boo also argue in favour of the slow rotation.  

Stokes $V$ data were collected in early 2008 only, with a rather dense coverage of the 
rotation cycle (see Table~\ref{tab:log569});  Stokes $V$ signatures are detected at all 
times, with longitudinal fields ranging from --50~G and --120~G and evolving smoothly 
with rotation phase.  The RV we measure ($-7.34\pm0.10$~\kms, with an internal dispersion 
of 0.03~\kms) agrees with that of \citet{Nidever02}.  
The rotational modulation of the Stokes $V$ profiles are reminiscent of those of AD~Leo 
(M08), suggesting that the star is not seen equator on;  the relative fluctuations of the 
longitudinal fields are however about twice larger than those of AD~Leo, indicating that 
$i$ is not as low as 20\degr\ (as for AD~Leo).  We chose $i=45\degr$ as a an intermediate 
value.  The rotational broadening in the spectral lines of CE~Boo is small, comparable to 
that of DS~Leo;  using the radius expected from theoretical models (0.43~\rsun, see
Table~\ref{tab:sample}), the rotation period of \citet{Kiraga07} and the inclination
angle we assumed ($i=45\degr$), we find (and used) $\vsini=1$~\kms.  
Conversely to the 2 previous stars, we have to assume $f=0.05$ (smaller than the usual 
value for mid-M dwarfs, see M08) to obtain a $\chinu=1$ to the Stokes $V$ profiles.  

%% fig3:
\begin{figure*}
\center{\includegraphics[scale=0.55,angle=-90]{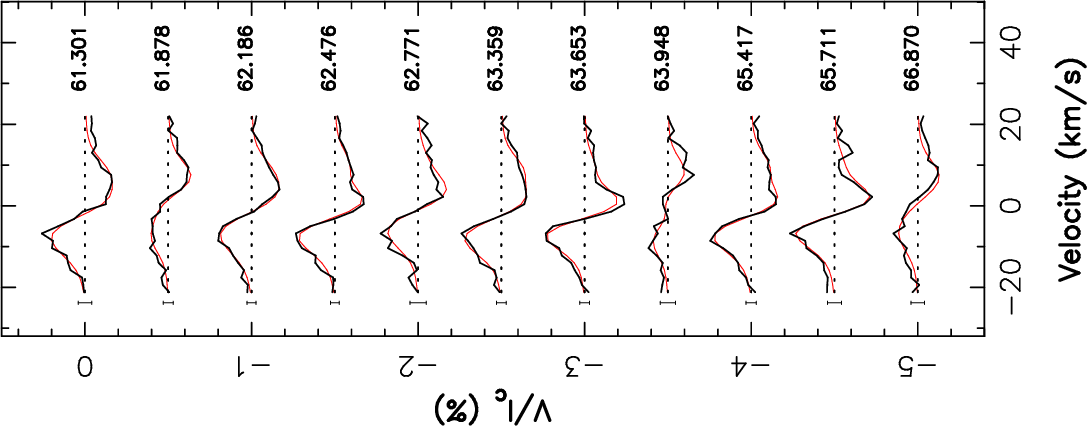}\hspace{1mm}
        \includegraphics[scale=0.55,angle=-90]{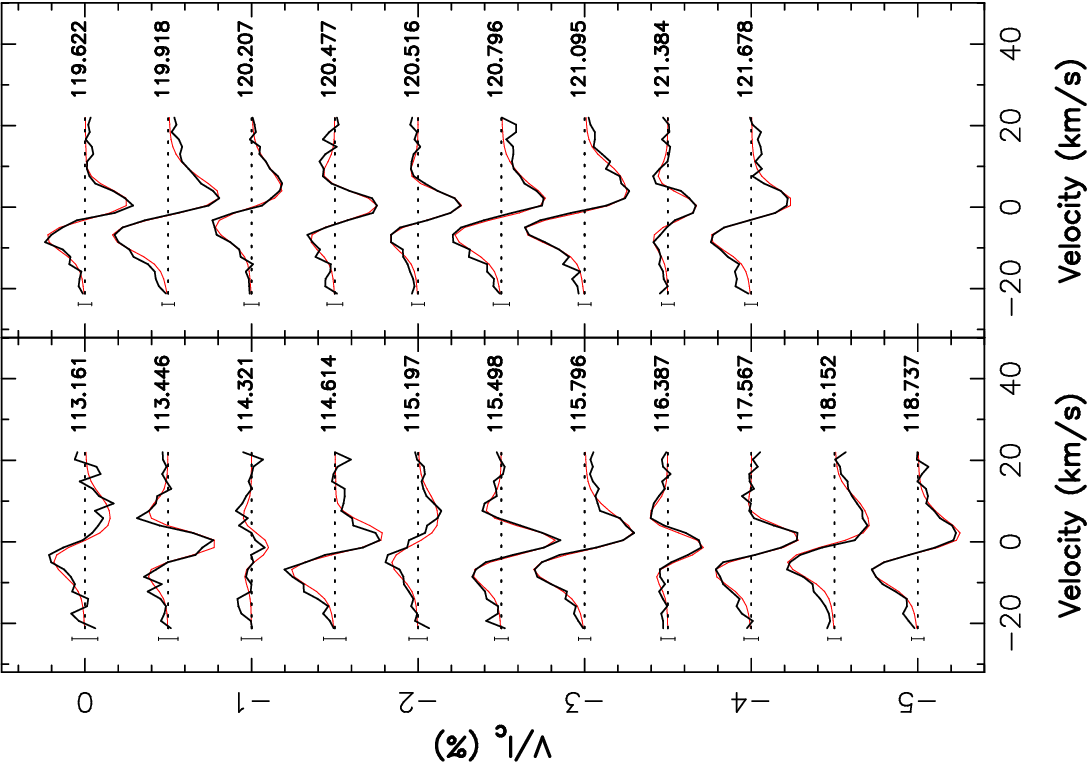}}
\caption[]{Same as Fig.~\ref{fig:494fit} for OT~Ser=GJ~9520.  } 
\label{fig:9520fit}
\end{figure*}

%% fig4:
\begin{figure*}
\center{\includegraphics[scale=0.65]{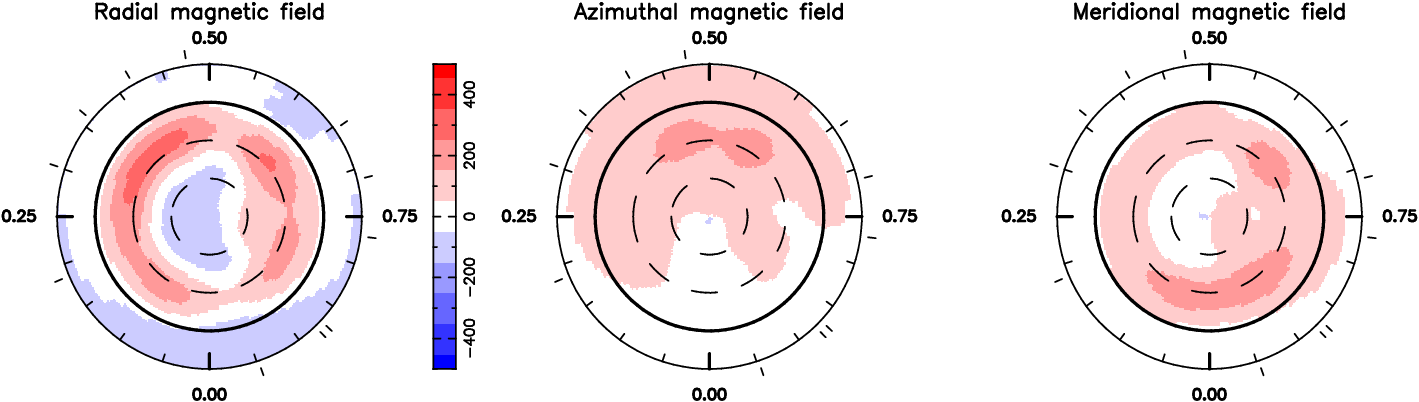}}
\center{\includegraphics[scale=0.65]{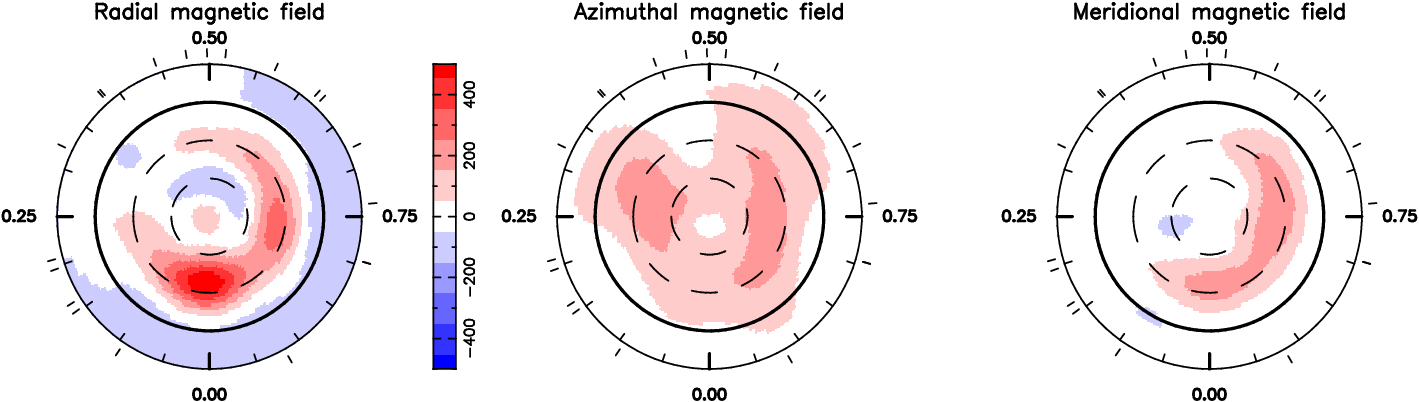}}
\caption[]{Same as Fig.~\ref{fig:494map} for OT~Ser=GJ~9520.  }
\label{fig:9520map}
\end{figure*}

The solid-body-rotation period providing the best fit to the Stokes $V$ data is equal to 
14.7~d, which we used to phase all our spectra.  This is slightly longer than the 
period found by \citet{Kiraga07}, suggesting that CE~Boo is also subject to differential 
rotation (as DT~Vir and DS~Leo) with at least $\dom=0.03$~\rpd\ (assuming our period 
yields the rotation rate at the pole and the period of \citealt{Kiraga07} traces the 
rotation rate at the equator).  Proceeding as above, we obtain no clear minimum in the 
\omeq--\dom\ domain, indicating that our data are not suitable for measuring differential 
rotation;  this is not too surprising given the fairly simple rotational modulation of the 
Stokes $V$ profiles and its moderate amplitude.  We therefore assumed that CE~Boo rotates 
as a solid-body in the following;  very similar results are obtained if assuming that 
CE~Boo is hosting differential rotation similar to that of DT~Vir and DS~Leo.  

The optimal maximum entropy fit to the Stokes $V$ data that we obtain is shown in 
Fig.~\ref{fig:569fit} and corresponds to $\chinu=1$, i.e., to a \chinu\ improvement over 
a non-magnetic model of $\times35$.  The reconstructed magnetic map is shown in Fig.~\ref{fig:569map}.   
The magnetic topology we derive is almost completely poloidal, with less that 10\% of 
the reconstructed energy concentrating into the toroidal component (see Table~\ref{tab:syn});  
the poloidal field 
is quite simple (93\% of the energy in the $\ell\leq2$ modes) and 
mostly axisymmetric (96\% of the energy in $m<\ell/2$ modes).

\section{OT Ser = GJ 9520 = HIP 75187}
\label{sec:gj9520}

OT~Ser is an active M1.5 dwarf with spectral lines showing significant rotational broadening.  
Located at an Hipparcos distance of $11.39\pm0.25$~pc, it has no identified 
companion \citep[][Forveille private communication]{Daemgen07}.  
Using the mass-luminosity relations of \citet{Delfosse00}, we estimate a mass of 
$0.55\pm0.02$~\msun.  
Two discrepant rotation periods, both estimated from photometric variability, are reported 
in the literature;  while \citet{Norton07} find $\prot=3.38$~d, \citet{Kiraga07} obtain 
$\prot=0.37$~d.  Given the estimated radius of OT~Ser (about 0.5~\rsun, see Table~\ref{tab:sample}), 
the second period would imply an equatorial velocity of almost 70~\kms, much larger than the 
observed width of spectral lines;  the rotation period of \citet{Norton07} is thus much more 
likely to be the correct one.  

Stokes $V$ data were collected in 2007 and 2008, with a rather dense coverage of the 
rotation cycle in 2008 (see Table~\ref{tab:log9520});  Stokes $V$ signatures are detected in 
almost all spectra.  The longitudinal field variations with phase are very different at both 
epochs (with \Bl\ varying from 55 to 80~G in 2007 and from --10 to 110~G in 2008) demonstrating 
that the magnetic topology changed significantly on a timescale of only 0.5~yr.  
The RV we measure ($6.77\pm0.10$~\kms\ and $6.83\pm0.10$~\kms, with internal dispersions  
of 0.03~\kms and 0.04~\kms) are slightly different at both epochs, possibly reflecting  
the change in the magnetic topology.  Modelling Stokes $I$ line profiles yields 
$\vsini=6\pm1$~\kms.  From the expected radius and rotation period, we infer that the star 
is seen at an intermediate inclination angle;  we use $i=45\degr$ in the following.  
As for CE~Boo, we have to adjust $f$ to obtain a $\chinu=1$ fit to the Stokes $V$ profiles;  
we find that $f$ equals 0.05 and 0.10 in 2007 and 2008 respectively.  

The solid-body-rotation period providing the best fit to the data is equal to 3.40~d at both 
epochs;  we used it to phase all our data (see Table~\ref{tab:log9520}).  
This period is close to but slightly different than that of \citet{Norton07}, suggesting 
that OT~Ser is also differentially rotating.  Fitting our 2008 Stokes $V$ data further 
confirms that OT~Ser is not rotating as a solid body;  with the same procedure as above, 
we obtain that $\omeq=1.88\pm0.01$~\rpd\ and $\dom=0.12\pm0.02$~\rpd, i.e., that the 
rotation periods at the equator and the pole are respectively equal to 3.34~d and 3.57~d.  
The photospheric shear of OT~Ser is thus apparently even stronger than that of DT~Vir 
and DS~Leo, with the equator lapping the pole by one cycle every $52_{-7}^{+11}$~d.  

The optimal maximum entropy fit to the Stokes $V$ data that we obtain (including 
the effect of differential rotation) is shown in Fig.~\ref{fig:9520fit} and 
corresponds to $\chinu=1$, i.e., to a \chinu\ improvement over a non-magnetic 
model of $\times23$ and $\times19$ for 2007 and 2008 respectively.  The 
reconstructed magnetic maps are shown in Fig.~\ref{fig:9520map}.  Although both maps 
show a similar large-scale topology (e.g., same latitudinal dependence of field 
polarities for all components), differences are nevertheless obvious;  for instance, 
the ring of positive radial field encircling the star at mid latitudes shows a 
prominent blob at phase 0.0 in 2008 (causing the large-amplitude longitudinal-field 
modulation observed at this epoch).   

The magnetic topologies we derive are dominantly poloidal, with about 20--30\% of 
the reconstructed energy concentrating into the toroidal component;  the poloidal field 
is mostly axisymmetric and includes a significant dipole component at both epochs 
(see Table~\ref{tab:syn}).

\section{GJ 182 = HIP 23200}
\label{sec:gj182}

GJ~182 is a very young single M0.5 dwarf of the IC~2391 supercluster located at an Hipparcos distance 
of $26.7\pm1.7$~pc.  The star is surrounded by a massive debris disk indicating on-going planetary 
formation \citep{Liu04}, further demonstrating that it is indeed very young.  
Its position in the HR diagram (about 0.5~mag above the main sequence) is in agreement with the age 
of its young moving group (about 35~Myr, e.g., \citealt{Montes01}).  
Using the evolutionary models of \citet{Baraffe98} and matching them to an absolute 
$V$ magnitude and a logarithmic luminosity (relative to the Sun) of $7.94$ and $-0.83$ 
respectively, we find that GJ~182 has a mass of $0.75$~\msun, a radius of $0.82$~\rstar, a 
temperature of 3950~K and an age of 25~Myr;  this is what we assume in the following.  
The high lithium content of GJ~182 suggests that the star is even younger, possibly as young as 
10--15~Myr \citep{Favata98} as evolutionary models predict that lithium should be already 
strongly depleted at 20~Myr \citep{Favata98}.  Effects of rotation and magnetic fields on the 
stellar structure and on the evolution \citep[e.g.,][not taken into account in
existing studies]{Chabrier07} are however likely to affect model predictions significantly.   

Stokes $V$ data were collected in 2007, covering only about half the rotation cycle of 
GJ~182 (see Table~\ref{tab:log182});  Stokes $V$ signatures are detected in all spectra and 
longitudinal field vary from --90~G to 20~G with rotation phase.  
The RV we measure ($19.35\pm0.10$~\kms, with an internal dispersion of 0.18~\kms) varies 
with rotational phase and correlate well with longitudinal field values;  
although the statistics is moderately significant (only 7 data points available), 
it suggests that the RV fluctuations we detect (full amplitude of about 0.4~\kms) are 
related to surface magnetic activity.  Spectral lines are significantly broadened by 
rotation;  modelling Stokes $I$ LSD profiles yield $\vsini=10\pm1$~\kms.
Photometric modulation indicates a rotation period of about 4.4~d \citep{Kiraga07},  
suggesting that the star is viewed equator-on rather than pole-on;  we therefore set the 
inclination angle at $i=60\degr$.  

The solid-body-rotation period providing the best fit to the data is equal to 4.35~d, 
slightly smaller than the photometric period of 4.41~d measured by \citet{Kiraga07};  
we used our estimate to phase all spectra, and take this as a likely indication that GJ~182 
is a differential rotator.  Despite the small number of spectra and the limited phase 
coverage, fitting our Stokes $V$ data down to $\chinu=1$ suggests that GJ~182 is indeed 
not rotating as a solid body.  Using the procedure described in Sec.~\ref{sec:mod},  
we obtain that $\omeq=1.46\pm0.01$~\rpd\ and $\dom=0.06\pm0.03$~\rpd;  the corresponding 
rotation periods at the equator and the pole are respectively equal to 4.30~d and 4.49~d 
(bracketing both our rotation period and that of \citealt{Kiraga07}).  Solid-body rotation 
is excluded at the 2$\sigma$ level;  more data are needed to confirm this with better 
precision.  

%% fig3:
\begin{figure}
\center{\includegraphics[scale=0.55,angle=-90]{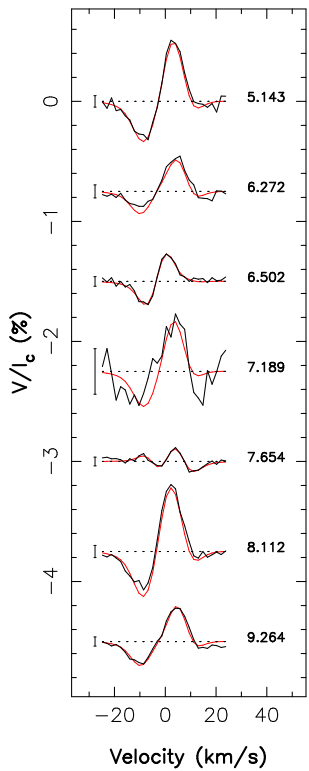}}
\caption[]{Same as Fig.~\ref{fig:494fit} for GJ~182.  }
\label{fig:182fit}
\end{figure}

%% fig4:
\begin{figure*}
\center{\includegraphics[scale=0.65]{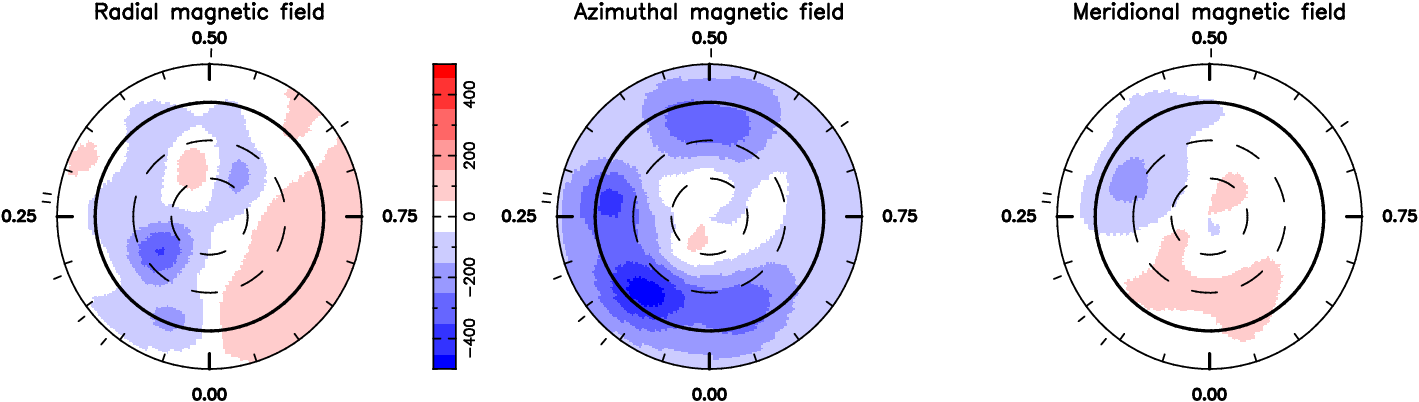}}
\caption[]{Same as Fig.~\ref{fig:494map} for GJ~182.  }
\label{fig:182map}
\end{figure*}

The optimal maximum entropy fit to the Stokes $V$ data that we obtain (including
the effect of differential rotation) is shown in Fig.~\ref{fig:182fit} and
corresponds to $\chinu=1$, i.e., to a \chinu\ improvement over a non-magnetic
model of $\times22$.  The reconstructed magnetic map is shown in Fig.~\ref{fig:182map}.  
The magnetic field is dominantly toroidal, and the poloidal component is mostly 
non-axisymmetric (see Table~\ref{tab:syn}).

\section{GJ 49 = HIP 4872}
\label{sec:gj49}

GJ~49 is a single M1.5 dwarf with sharp spectral lines and relatively low activity;  
it is the least active star of our sample (see Table~\ref{tab:sample}).  The Hipparcos 
distance is equal to $10.06\pm0.14$~pc.  
No rotation periods are found in the literature;  the RV is reported to be 
$-5.97\pm0.10$ \citep{Nidever02}.  Using the mass-luminosity relations of 
\citet{Delfosse00}, we estimate a mass of $0.57\pm0.02$~\msun.  

Stokes $V$ data were collected in 2007, covering the whole rotation cycle of GJ~49 
(see Table~\ref{tab:log49});  Stokes $V$ signatures are detected in all spectra, with 
longitudinal fields ranging from --30~G to 0~G across the cycle.  
The RV we measure ($-6.02\pm0.10$~\kms, with an internal dispersion of 0.02~\kms) is 
in good agreement with that of \citet{Nidever02}.  Modelling Stokes $I$ LSD profiles 
indicates that the rotational broadening is very small;  we thus set $\vsini=1$~\kms.  

We determine the rotation period by selecting the one with which the Stokes $V$ profiles 
can be fitted at $\chinu=1$ with smallest magnetic energy in the reconstructed image;  
we find that $\prot=18.6\pm0.3$~d;  we also find that an intermediate inclination 
angle ($i\simeq45\degr$) minimises the amount of reconstructed magnetic information.  
We find that the data are compatible with solid-body rotation, but the accuracy to which 
we measure \dom\ (error bar $\simeq0.05$~\rpd) is not high enough to know whether 
GJ~49 also hosts differential rotation similar to that found on the other sample stars.  

The fit to the Stokes $V$ data that we obtain is shown in Fig.~\ref{fig:49fit} and
corresponds to a \chinu\ improvement over a non-magnetic
model of $\times11$.  The reconstructed magnetic map is shown in Fig.~\ref{fig:49map}.  
The poloidal and toroidal field components roughly share the same amount of energy, 
with the poloidal field being mainly dipolar and axisymmetric (see Table~\ref{tab:syn}).

\section{Summary and discussion}
\label{sec:disc}

We report in this paper the results of our spectroscopic survey of M dwarfs;  
following M08 (concentrating on mid-M dwarfs), we describe here the Zeeman signatures 
and the large-scale magnetic topologies we observed on 6 early-M dwarfs (from M0 to M3).  
We also determined or confirmed the rotation period of all stars (ranging from 2.8 to 
18.6~d), and detected significant surface differential rotation in 4 of them (with a 
strength comparable to that of the Sun).  

%% fig3:
\begin{figure}
\center{\includegraphics[scale=0.55,angle=-90]{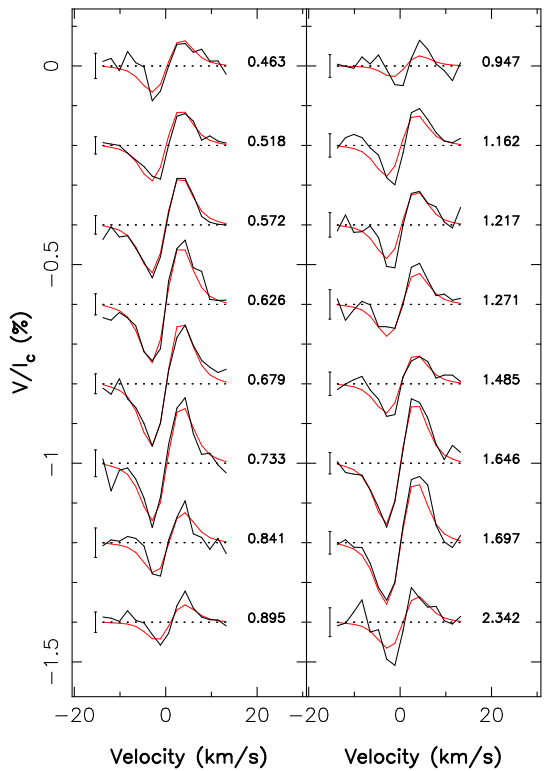}}
\caption[]{Same as Fig.~\ref{fig:494fit} for GJ~49.  }
\label{fig:49fit}
\end{figure}

%% fig4:
\begin{figure*}
\center{\includegraphics[scale=0.65]{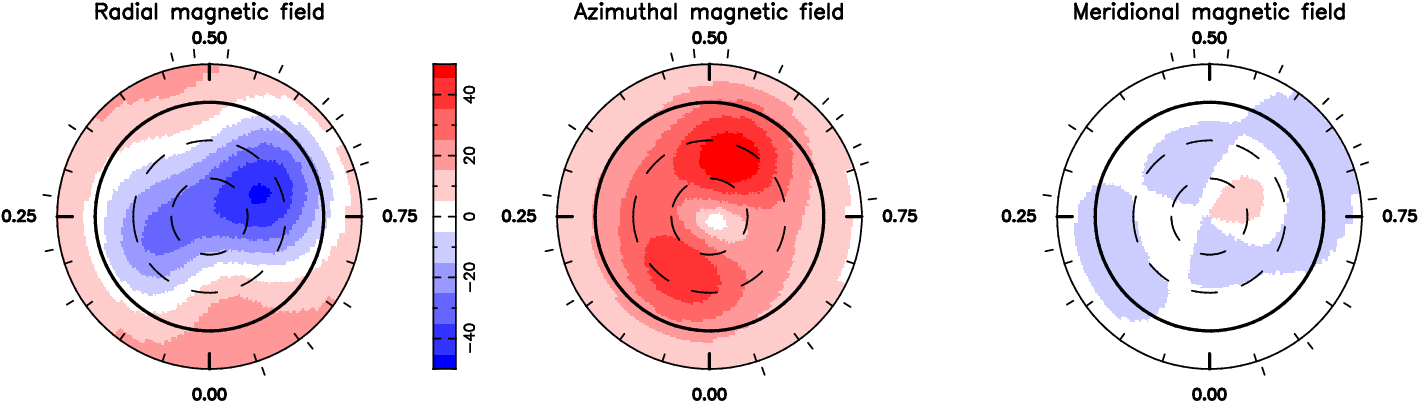}}
\caption[]{Same as Fig.~\ref{fig:494map} for GJ~49.  }
\label{fig:49map}
\end{figure*}

The magnetic fields we detect in early-M dwarfs are weak (typically a 
few tens of G), smaller in particular than those found in mid-M dwarfs (M08) 
by typically a factor of 5 (see Fig.~\ref{fig:diag}) with a sharp transition 
occuring at 0.4~\msun\ (see Fig.~\ref{fig:ross}, left panel).  The large-scale 
magnetic topologies we derive are also significantly different, involving a 
much larger fraction of toroidal fields and a lower axisymmetric degree of 
poloidal fields whenever $\mstar>0.5$~\msun\ (5 stars in the present sample);  
below 0.5~\msun, the poloidal field is largely dominant and axisymmetric and 
its strength is increasing rapidly as mass decreases.  We also observe that 
the typical lifetime of the large-scale magnetic topology is very different 
on both sides of the 0.4--0.5~\msun\ threshold, with lifetimes smaller than a few 
months on the hot side and longer than 1~yr on the cool side (M08).  
This threshold is very sharp and well defined, with little apparent dependence 
with the rotation period.  

At this stage, it is interesting to consider the effective Rossby number $Ro$, 
defined as $Ro=\prot/\tau_{\rm c}$ where $\tau_{\rm c}$ is the convective 
turnover time\footnote{The $\tau_{\rm c}$ values that we use here are those of 
\citet{Kiraga07}, determined empirically from relative X-ray luminosities 
of stars with different masses and rotation periods.  At masses of 
$\simeq1$~\msun, they match the usual value of $\simeq15$~d;  they steeply 
increase with decreasing mass below masses of $\simeq0.6$~\msun.};  in 
particular, $Ro$ is a convenient parameter for comparing the strength of 
dynamo action (and, e.g., \logRx\ that indirectly reflects dynamo action through 
coronal heating) in stars with different masses.  Figure~\ref{fig:ross} (right 
panel) illustrates how \logRx\ smoothly varies with $Ro$ for both early- and mid-M 
dwarfs studied here and in M08;  we find that \logRx\ increases steeply with 
decreasing $Ro$ until $Ro\simeq0.1$ where \logRx\ saturates at a level of 
about --3.0, in good agreement with previous studies \citep[e.g.,][]{James00}.  
Only GJ~182 (at $Ro=0.17$) lies slightly above the overall trend, as a likely 
consequence of its extreme youth and the related differences in its internal 
structure.  

\begin{figure*}
\includegraphics[scale=0.6,angle=-90]{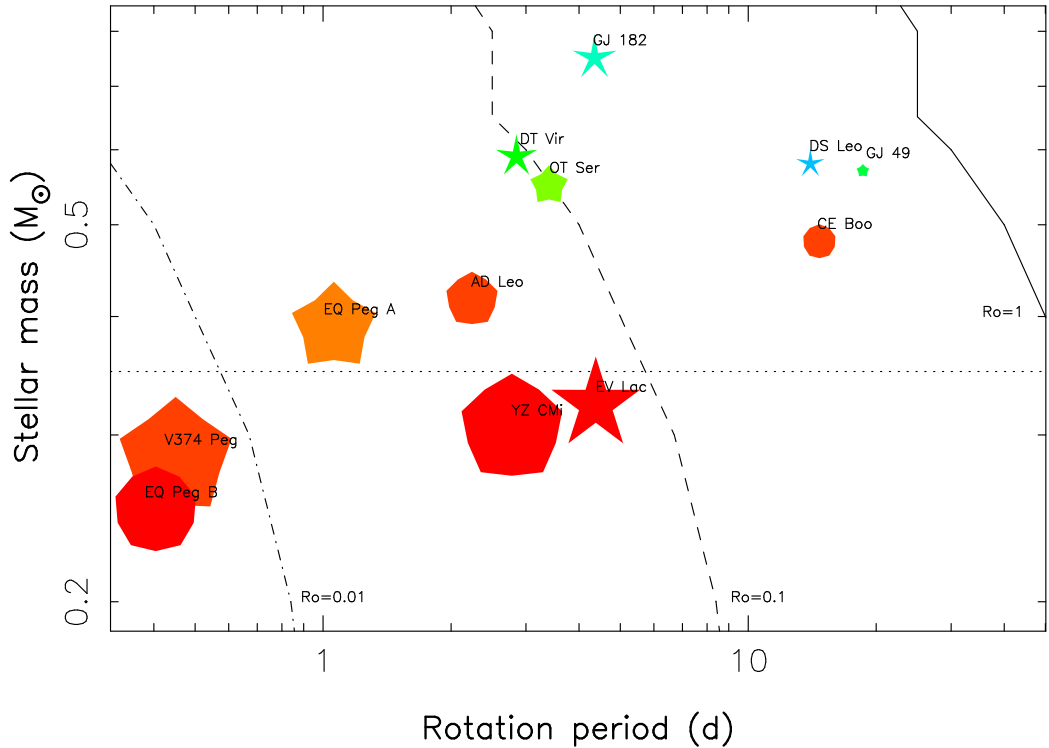} 
\caption[]{Basic properties of the large-scale magnetic 
properties of early- and mid-M dwarfs as a function of stellar mass 
and rotation period.  Symbols size indicates magnetic energies, symbol 
colour illustrates the field configuration (blue and red for purely 
toroidal and purely poloidal fields respectively) while symbol shape 
depicts the degree of axisymmetry of the poloidal field component 
(decagon and stars for a purely axisymmetric and purely non-axisymmetric 
poloidal fields respectively).  Results for early-M stars are from 
this paper and results for mid-M stars are from M08.  The full, dashed and 
dash-dot lines respectively trace the location of the $Ro=1$, 0.1 and 0.01 
contours, while the dotted line shows the theoretical full-convection threshold 
($\mstar\simeq0.35$~\msun).}
\label{fig:diag}
\end{figure*}

%% fig3:
\begin{figure*}
\center{\includegraphics[scale=0.37,angle=-90]{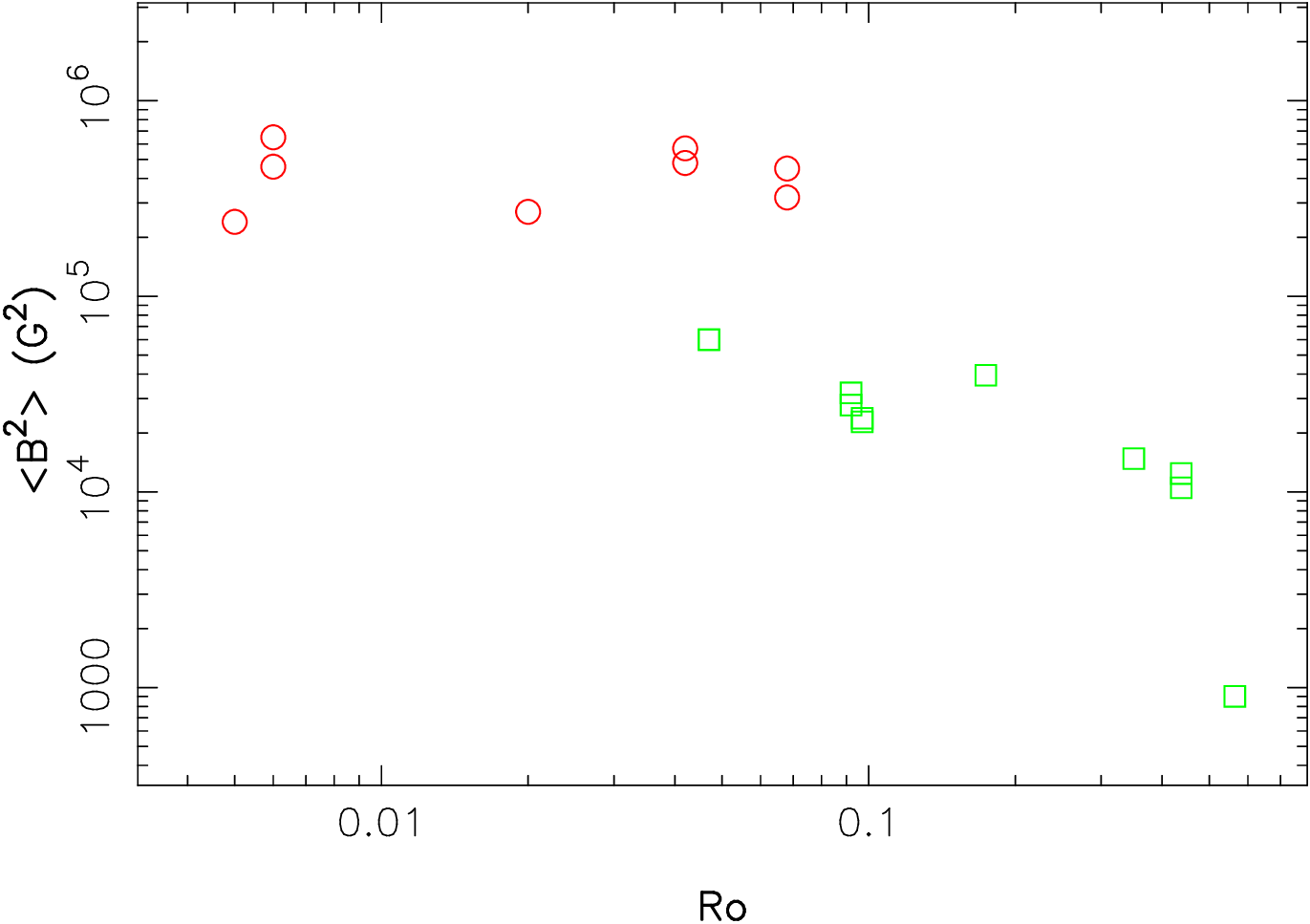}\hspace{5mm}
        \includegraphics[scale=0.37,angle=-90]{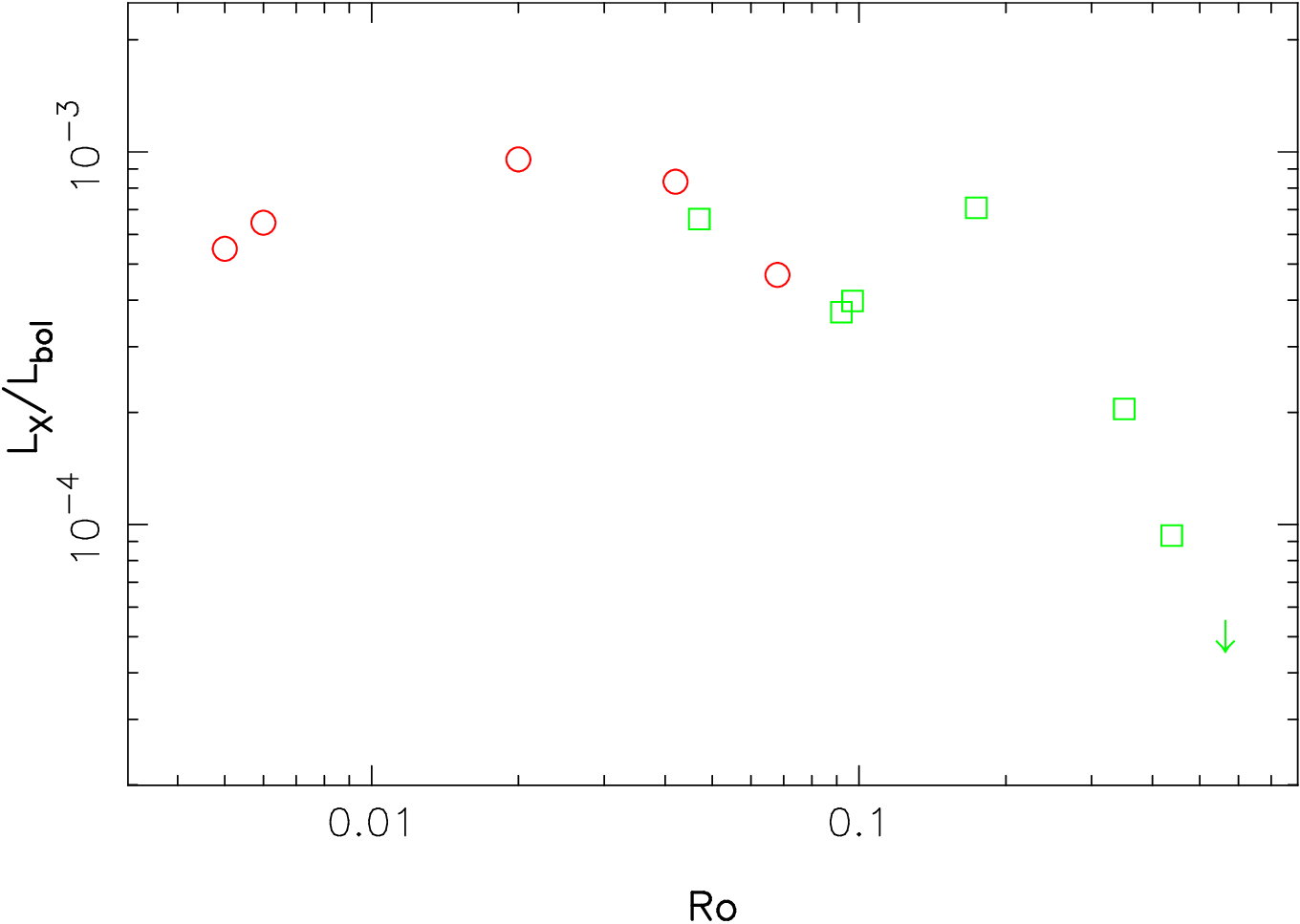}}
\caption[]{Reconstructed magnetic energy (left) and relative X-ray luminosity (with 
respect to the bolometric luminosity, right) as a function of $Ro$ for stars studied in 
this paper and in M08.  Stars with masses larger and smaller than 0.4~\msun\ are shown 
as green squares and red circles respectively.  In the left panel, measurements at 
different epochs (whenever available) are shown for each star to illustrate the 
typical scatter expected from temporal variability.   }
\label{fig:ross}
\end{figure*}

We note that the abrupt step in the large-scale magnetic energy between early- 
and mid-M dwarfs (see Fig.~\ref{fig:ross}, left panel) apparently correlates better 
with stellar mass than with $Ro$;  at $Ro=0.05$, AD~Leo exhibits a magnetic flux 
compatible with that of all other early M dwarfs, but significantly weaker than 
that of EV~Lac (at $Ro=0.07$) and YZ~CMi (at $Ro=0.04$).  More data are needed to 
confirm this point, especially for high-mass rapid rotators (i.e., having 
$Ro\simeq0.01$ and $\mstar>0.5$~\msun) and low-mass slow rotators (with $Ro>0.1$ 
and $\mstar<0.4$~\msun).  We also note that this 
abrupt step does not show up in the \logRx\ vs $Ro$ plot of Fig.~\ref{fig:ross} 
(right panel); this is presumably because X-rays are sensitive to overall 
magnetic energies while we are only sensitive to the largest scales.  Our result 
therefore suggests that, at some specific stellar mass ($\simeq0.4$~\msun, rather 
than at some specific $Ro$), dynamo processes become suddenly much more efficient 
at triggering large-scale magnetic fields;  we also observe that, at more or less 
the same mass ($\simeq0.5$~\msun), large-scale topologies of M dwarfs become 
dominantly poloidal and axisymmetric.  Note however that, even in the case of 
mid-M dwarfs, the large-scale fields we derive are significantly smaller than 
the corresponding equipartition field (a few kG, M08).  

\begin{table*}
\caption{Properties of the large-scale field topologies derived in 
the present study.  For each star, observations at different epochs 
are listed separately.  The table lists sequentially the name of the star, 
the mass, the rotation period, the effective Rossby number and the logarithmic 
relative X-ray luminosity (taken from Table~\ref{tab:sample}), the surface 
angular rotation shear between the equator and pole \dom\ (whenever detected), 
the derived filling factor $f$ (whenever applicable), the reconstructed magnetic 
energy and flux (i.e., $<B^2>$ and $<B>$), and the fractional energies in the 
poloidal field, the poloidal dipole ($\ell=1$) modes, the poloidal quadrupole 
($\ell=2$) modes, the poloidal octupole ($\ell=3$) modes, and the poloidal 
axisymmetric ($m<\ell/2$) modes.  All field components with $\ell\leq3$ 
are required to fit the data at noise level.  }
 \begin{tabular}{rccccccccccccc} 
\hline
Star  & \mstar & \prot & $Ro$ & $\logRx$ & \dom & $f$ & $<B^2>$ & $<B>$ & pol & dip & qua & oct & axi \\
      & (\msun) & (d)  &      &            &  (\rpd) &  & ($10^4$~G$^2$) & (G) &  &  &  &  & \\
\hline
GJ 182 2007 & 0.75 & 4.35 & 0.174 & --3.1 & $0.06\pm0.03$   &      & 3.95 & 172 & 0.32 & 0.48 & 0.18 & 0.14 & 0.17 \\ 
DT Vir 2007 & 0.59 & 2.85 & 0.092 & --3.4 &                 &      & 2.78 & 145 & 0.38 & 0.64 & 0.17 & 0.08 & 0.12 \\  
       2008 &      &      &       &       & $0.060\pm0.006$ &      & 3.21 & 149 & 0.53 & 0.10 & 0.17 & 0.17 & 0.20 \\ 
DS Leo 2007 & 0.58 & 14.0 & 0.438 & --4.0 &                 &      & 1.24 & 101 & 0.18 & 0.52 & 0.37 & 0.08 & 0.58 \\ 
       2008 &      &      &       &       & $0.076\pm0.020$ &      & 1.05 &  87 & 0.20 & 0.52 & 0.31 & 0.07 & 0.16 \\ 
GJ 49  2007 & 0.57 & 18.6 & 0.564 &$<-4.3$&                 &      & 0.09 &  27 & 0.48 & 0.71 & 0.20 & 0.07 & 0.67 \\ 
OT Ser 2007 & 0.55 & 3.40 & 0.097 & --3.4 &                 & 0.05 & 2.28 & 136 & 0.80 & 0.47 & 0.19 & 0.18 & 0.86 \\ 
       2008 &      &      &       &       & $0.12\pm0.02$   & 0.10 & 2.38 & 123 & 0.67 & 0.33 & 0.17 & 0.21 & 0.66 \\ 
CE Boo 2008 & 0.48 & 14.7 & 0.350 & --3.7 &                 & 0.05 & 1.48 & 103 & 0.95 & 0.87 & 0.06 & 0.03 & 0.96 \\ 
\hline
 \end{tabular} 
\label{tab:syn} 
\end{table*}

Significant surface toroidal fields are detected even in DS~Leo and GJ~49, i.e., 
the two slowest rotators with masses larger than 0.5~\msun;  it suggests that the 
transition between mainly poloidal and mainly toroidal fields in $\mstar>0.5$~\msun\ 
stars occurs at $Ro\simeq0.5-1.0$, with the Sun located on the other side of this 
boundary (at $Ro\simeq1.5-2.0$).  Note that this boundary coincides with the sharp 
onset of photometric variability in convective stars (occuring below $Ro\simeq0.7$, 
\citealt{Hall91}).  
With a poloidal field concentrating 70--80\% of the reconstructed 
magnetic energy, OT~Ser is off this trend;  we suspect that this is due to 
its proximity with the 0.5~\msun\ sharp threshold below which magnetic 
topologies become dominantly poloidal.  

Early-M dwarfs are found to show significant differential rotation;  the 
values we obtain for the surface angular rotation shear \dom\ ranges from 0.06 
to 0.12~\rpd, i.e., from once to twice the strength of the surface latitudinal 
shear of the Sun.  Our detection is further confirmed by the small (but significant) 
differences between the rotation periods we measure and the values reported in 
the literature (derived from photometric fluctuations) and by the short lifetimes 
of the large-scale field topologies (quickly distorted beyond recognition by 
differential rotation).  
Previous Doppler imaging studies of early-M dwarfs with very fast rotation 
($Ro\simeq0.01$) report that differential rotation is very 
small \citep{Barnes05};  our study suggests that the situation may significantly 
differ in moderate rotators like those we considered.  
Our result is also different from what is observed in mid-M dwarfs where 
differential rotation is very small (a few \mrpd\ at most, i.e., more than 10 times 
smaller than that of early-M dwarfs, M08) and large-scale magnetic topologies 
long-lived (M08).  It is not clear yet what this difference is due to;  while 
small $Ro$ may contribute at freezing differential rotation, this is likely not 
the only relevant parameter for this problem (e.g., with DT~Vir and EV~Lac showing 
respectively significant and no differential rotation despite their similar $Ro$).  

The sharp transition that we report between the magnetic (and differential rotation) 
properties of early- and mid-M dwarfs is surprising at first glance;  naively, one 
would expect the properties of large-scale magnetic fields to change smoothly with 
stellar mass as the radiative core gets progressively smaller.  From the evolutionary 
models of \citet{Siess00}, we however note that the outer radius of the radiative 
core of early-M dwarfs is changing very quickly with stellar mass, from about 
0.5~\rstar\ for a 0.5~\msun\ star to a negligible fraction for a 0.4~\msun\ star.  
We speculate that this sharp transition is the main reason for the abrupt magnetic 
threshold that we report here.  The rapid increase in empirical convective turnover 
times occuring at about the same location \citep{Kiraga07} also likely contributes 
at making the transition between both dynamo regimes very sharp.  

The most recent 
dynamo simulations of fully convective M dwarfs \citep{Browning08} (carried out 
for $Ro\simeq0.01$) are successful at reproducing the frozen differential rotation 
that we observe (M08);  they however predict the presence of strong toroidal fields 
that we do not see in mid-M dwarfs with similar $Ro$.  
We speculate that the abrupt change in the large-scale magnetic topology of M dwarfs 
that we report here to occur at spectral type M3 may also be (at least partly) 
responsible for the reduced magnetic braking observed for stars later than M3 
\citep[e.g.,][]{Delfosse98};  MHD simulations of magnetic winds are necessary to 
estimate quantitatively whether the observed change in the large-scale magnetic 
topology can indeed explain the longer spin-down timescales.  

In the two stars having \vsini\ measured with sufficient precision (i.e., $\vsini\geq10$, 
GJ~182 and DT~Vir), we find that $\rstar \sin i$ (equal to $0.86\pm0.09$~\rsun\ and 
$0.62\pm0.06$~\rsun\ respectively) is already larger than the predicted radius from 
theoretical models \citep{Baraffe98};  while this could result from overestimating the 
true age (and hence underestimating the true radius) of GJ~182, this explanation 
does not apply for DT~Vir, for which we conclude that the observed radius is truly larger 
(by at least 10\% and potentially as much as 30\%) that what theoretical models predict.  
A similar conclusion is reached for V374~Peg \citep{Donati06, Morin08};  furthermore, 
M08 obtains that $\rstar\sin i$ is equal to the predicted theoretical radius (within the 
error bars) for 4 other active stars, suggesting again that \rstar\ is larger than expected.  
Following \citet{Chabrier07}, we propose that this effect is a direct consequence of magnetic 
fields getting strong enough (and hence saturating the dynamo, see Fig.~\ref{fig:ross}) 
to affect the energy transport throughout the convective zone and hence the radius.  
Our results therefore independently confirm the report that cool low-mass active stars, 
either single \citep{Morales08} or within close eclipsing binaries \citep{Ribas06}, 
usually have oversized radii with respect to inactive stars of similar spectral types.  

We also detect significant RV fluctuations (with a full amplitude of up to 0.40~\kms) 
in the 3 very active stars of our sample (with rotation periods smaller than 5~d).  For 
the most active ones (DT~Vir and GJ~182, showing the largest RV modulation), the RV 
variations correlate reasonably well (though not perfectly) with longitudinal fields, 
suggesting that the origin of the variations is indeed the magnetic field (and the 
underlying activity).  It also suggests that spectropolarimetric observations 
should be carried out simultaneously with RV measurements of active stars to enable 
filtering out efficiently the activity jitter from the RV signal;  this technique may 
prove especially useful when looking at Earth-like habitable planets orbiting around 
M dwarfs in the future, e.g., with a nIR spectropolarimeter such as SPIRou (a nIR 
counterpart of ESPaDOnS, proposed for CFHT).  

Our spectropolarimetric survey is an on-going study;  we are now concentrating 
on late-M dwarfs (M5-M8) to derive similar observational constraints about the 
large-scale magnetic topologies of stars in the yet unexplored 
0.08--0.20~\msun\ region of Fig.~\ref{fig:diag} to investigate how dynamo 
processes operate down to the brown dwarf threshold, i.e., when stellar 
atmospheres get so cool that they start to decouple from their magnetic fields.  

\section*{Acknowledgements}

We thank the TBL staff for their help during data collection.  We also
thank the referee, J.D.~Landstreet, for valuable comments on the manuscript, as 
well as G.~Chabrier, J.~Bouvier and M.~Browning for enlightening discussions on 
various topics discussed in this paper.

%% Bibliography

\bibliography{mdwarf2}

\bibliographystyle{mn2e}

\end{document}